\documentclass[%
reprint,
superscriptaddress,
preprintnumbers,
 amsmath,amssymb,
 aps,
prb,
]{revtex4-2}

\usepackage{subfiles}
\usepackage{graphicx}
\usepackage{dcolumn}
\usepackage{bm}
\usepackage{hyperref}
\usepackage{cleveref} 

\usepackage[version=4]{mhchem}
\usepackage[final,numbered]{matlab-prettifier}
\usepackage{lineno}
\usepackage{booktabs} 

\begin{document}

\author{Ang Feng}
 \affiliation{LumiLab, Department of Solid State Sciences, Faculty of Sciences, Ghent University, Krijgslaan 281-S1, Ghent 9000, Belgium}
 \affiliation{Center for Nano- and Biophotonics (NB-Photonics), Ghent University,  Ghent 9000, Belgium}
 \author{Jonas J. Joos}
   \email{Jonas.Joos@UGent.be}
 \affiliation{LumiLab, Department of Solid State Sciences, Faculty of Sciences, Ghent University, Krijgslaan 281-S1, Ghent 9000, Belgium}
 \affiliation{Center for Nano- and Biophotonics (NB-Photonics), Ghent University,  Ghent 9000, Belgium}
\author{Jiaren Du}
 \affiliation{LumiLab, Department of Solid State Sciences, Faculty of Sciences, Ghent University, Krijgslaan 281-S1, Ghent 9000, Belgium}
 \affiliation{Center for Nano- and Biophotonics (NB-Photonics), Ghent University,  Ghent 9000, Belgium}
  \affiliation{International Joint Research Center for Photo-responsive Molecules and Materials, School of Chemical and Material Engineering, Jiangnan University, Wuxi 214122, China}
\author{Philippe F. Smet}%
  \email{Philippe.Smet@UGent.be}
 \affiliation{LumiLab, Department of Solid State Sciences, Faculty of Sciences, Ghent University, Krijgslaan 281-S1, Ghent 9000, Belgium}
 \affiliation{Center for Nano- and Biophotonics (NB-Photonics), Ghent University,  Ghent 9000, Belgium}

\title{Revealing trap depth distributions in persistent phosphors with a thermal barrier for charging}
\thanks{This is the postprint of the article published as \href{https://doi.org/10.1103/PhysRevB.105.205101}{Physical Review B 105, 205101 (2022).}}%

%

\date{\today}

\begin{abstract}
The performance of persistent phosphors under given charging and working conditions is determined by the properties of the traps that are responsible for these unique properties. Traps are characterized by the height of their associated barrier for thermal detrapping, and a continuous distribution of trap depths is often found in real materials. Accurately determining trap depth distributions is hence of importance for the understanding and development of persistent phosphors. However, extracting the trap depth distribution is often hindered by the presence of a thermal barrier for charging as well, which causes a temperature-dependent filling of traps. For this case, we propose a method for extracting the trap depth distribution from a set of thermoluminescence (TL) curves obtained at different charging temperatures. The TL curves are first transformed into electron population functions via the Tikhonov regularization, assuming first-order kinetics. Subsequently, the occupation of the traps as a function of their depth, quantified by the so-called \textit{filling function}, is obtained. Finally, the underlying trap depth distribution is reconstructed from the filling functions. The proposed method provides a substantial improvement in precision and resolution for the trap depth distribution compared with existing methods. This is hence a step forward in understanding the (de)trapping behavior of persistent and storage phosphors.
\end{abstract}

\maketitle


\section{Introduction}
Ideally, properties of materials can be tuned by tweaking only a few intrinsic parameters of these materials. In the case of persistent phosphors, the key property is the luminescence afterglow that can range from seconds to days after stopping the optical excitation, often called \textit{charging} \cite{Koen_EuperL_rev,Koen_nonEuperL_rev,JXuPerLRev,DPoelmanJAP}. High persistent luminescence (PersL) intensity and long PersL duration are two desirable properties under given charging and working conditions \cite{SMET2015_PersL}. One critical parameter controlling these properties is the density of traps, i.e., the absolute number of active traps per unit volume of the persistent phosphor \cite{David_capacity}. The higher the trap density, the more electrons a phosphor can store 
at the given charging condition, enhancing PersL intensity. Electrons are considered common charge carriers, although holes can act as charge carriers in certain cases \cite{LyuCEJ,JXuPerLRev}. Another parameter is the density of traps as a function of their depth $E_t$, which quantifies the energy barrier that trapped electrons must overcome thermally to recombine with holes. It is usually called \textit{trap depth distribution} for short, being denoted as $N(E_t)$. These parameters are scientifically important. On the one hand, they are useful to understand and thus to tailor performance of persistent phosphors under different  conditions. For example, the trap depth distribution can be translated into TL curves, whose intensity shows a linear relationship with respect to the total PersL intensity for a given charging and working temperature \cite{jrdu_Topt,jrdu_LPRev}. 
On the other hand, such parameters act as intrinsic materials parameters that can be compared across different phosphors, enabling the discovery of empirical rules in persistent phosphors. 
However, there are multiple difficulties when it comes to extracting the trap density and trap depth distribution. After charging ($t=0$), the density of trapped electrons (called here the electron population function) $n(E_t,q,t)$ is determined by the trap depth distribution $N(E_t)$ and the filling function $f(E_t,\Delta E, q,t)$ via,
\begin{linenomath*}
\begin{equation}
n(E_t,q,t) = f(E_t,\Delta E,q,t)N(E_t)
\label{eqn_fepopNE}
\end{equation}
\end{linenomath*}
where the experimental charging settings are collected in $q = [I_e(\lambda), t_\text{ch}, T_\text{ch}]$, with the irradiance $I_e(\lambda)$, charging duration $t_\text{ch}$ and charging temperature $T_\text{ch}$. In the trap depth range $[E_t,E_t+dE_t]$, the filling function $f(E_t,\Delta E,q,t)$ indicates the fraction of the traps that are filled at time $t$ after charging with settings $q$. In Eq.~\ref{eqn_fepopNE}, $\Delta E$ is the thermal barrier for charging, whose existence causes the maximum of the filling function to be dependent on charging temperature (for a given set $I_e(\lambda)$ and $t_\text{ch}$). This phenomenon has been observed in many persistent phosphors, for example SrAl$_2$O$_4$:Eu$^{2+}$ \cite{jonasb_trapping_detrapping_SAO}, Sr$_2$MgSi$_2$O$_7$:Eu$^{2+}$ \cite{Tydtgat2016Optically}, $M_2$Si$_5$N$_8$:Eu ($M$ = Ca, Sr, Ba) \cite{T_PersLMSiN258}, Y$_3$Al$_5$O$_{12}$:Ce$^{3+}$ \cite{YAGCeTanabe_JPCC},Y$_3$Al$_{5-x}$Ga$_x$O$_{12}$:Pr$^{3+}$ \cite{YAGPr_Ueda} and other garnets \cite{GarnetConductivity_JAPTanabe,PCGarnet_PCCPTanabe,GaGaAlOGarnet_CeesRonda,Rev_UedaOptMaterX}. This thermal barrier for charging severely complicates the extraction of trap depth distributions from experiments.  

The first obstacle is to recover the electron population function $n(E_t,q,t)$ from experimental TL curves. In literature, methods have been proposed to approximate the trap depth distribution by the electron population function itself, assuming that traps can be fully filled at the given charging condition. The simplest ansatz for an electron population function is a delta distribution, $N_0\delta(E_t-E_o)$, characterized by a single trap depth, $E_o$, and the total number of traps per unit volume, $N_0$. The quantity $N_0$ can be obtained by the method proposed by Van der Heggen \textit{et al.} \citep{David_capacity}. The discrete trap depth $E_o$ can be estimated by several methods \cite{Etrap_ChenRevJAP,EtMethod_Rev}, such as the initial rise method \cite{e_trapmechanism_ProcPhysSoc, AccValid_IRM}, the Urbach relation \cite{Ubach1930Zur}, and the analysis of position and symmetry of the glow curve \cite{Etrap_ChenRevJAP,EtMethod_Rev}. Alternatively, more advanced methods have been proposed to infer $n(E_t,q,t)$ in case that a nontrivial trap depth distribution is assumed. For example, the fractional glow technique \cite{FractionalGlow1966,FracGlow_correct} and the $T_m\text{-}T_\mathrm{stop}$ method \cite{McKeever1980On,AccValid_IRM} approximate $n(E_t,q,t)$ by thermally releasing trapped electrons in certain trap depth ranges by specially designed heating procedures. Recently, Khanin \textit{et al.} recovered $n(E_t,q,t)$ directly from a TL curve by numerical regularization, assuming first-order kinetics \cite{InvsProbJPCA}.

The main difficulty of extracting the trap depth distribution $N(E_t)$ lies in approximating the filling function after charging $f(E_t,\Delta E,q,t)$, especially when there exists a thermal barrier for charging, i.e., $\Delta E \neq 0$. At a given charging temperature $T_\text{ch}$, $f(E_t,\Delta E,q,t=0)$  approaches the Fermi-Dirac function in the limit of $t_\text{ch}\rightarrow\infty$ \cite{braunlich_ThermalRelax},
\begin{linenomath*}
\begin{equation}
f(E_t ,\Delta E,q,t=0)=\frac{f_0(\Delta E,q)}{1+\textrm{exp}\left(-\frac{E_t-E_f}{k_{\textsc{b}}T_\text{ch}}\right)} \label{eqn_fepop_f0}
\end{equation}
\end{linenomath*}
where $k_\textsc{b}$ is the Boltzmann constant, $E_{f}$ the quasi-Fermi level and $f_0(\Delta E,q)$ the magnitude of the filling function. $f_0(\Delta E,q)$ is <1.0 because of various detrapping routes, such as thermal detrapping and optically stimulated detrapping \cite{Tydtgat2016Optically,DavidIQE}. To access a wide range of $E_f$, phosphors are often charged at variable temperature $T_\text{ch}$ with fixed irradiance $I_{e}$ and fixed duration $t_\text{ch}$.  When a thermal barrier for charging $\Delta E$ is absent, $f_0(\Delta E,q)$ is independent of $T_\text{ch}$. The trap depth distribution in a range of $[E_f(T_\text{ch}),E_f(T_\text{ch}+\Delta T_\text{ch})]$ can be approximated by the difference of the total number of trapped electrons (per unit volume) \cite{Koen_trap_depth}. Experimentally, this can be determined from the difference in the integrated intensity of TL glow curves obtained at variable charging temperature $T_\text{ch}$. However, the thermal barrier $\Delta E$ poses two challenges for extracting the trap depth distribution $N(E_t)$. One is to approximate $f_0(\Delta E,q)$ for each filling function. The other is to reconstruct the trap depth distribution $N(E_t)$ from various individual pairs of $f(E_t,\Delta E,q,t)$ and $n(E_t,q,t)$. It is noteworthy that the thermal barrier for charging $\Delta E$ is treated as an empirical parameter to account for the temperature dependence of $f_0(\Delta E,q)$. This barrier is typically different from the activation energy of thermal quenching (TQ), which often originates from the crossover of potential energy curves due to electron-phonon interaction \cite{curie_luminescence_1960} or the thermal ionization of electrons at the excited states of the involved luminescent activators \cite{Dorenbos_TQ,deltaE_MNikl}.  

In this paper, we propose and validate a method, relying on first order kinetics, that circumvents the influence of $\Delta E$ and allows us to extract the trap depth distribution $N(E_t)$ from TL experiments. The phosphor BaSi$_2$O$_2$N$_2$:2\%Eu$^{2+}$ is used here as case study. This material shows desirable PersL \cite{PersLBaSiON} and mechanoluminescence \cite{MLRevMaterials,MLinBaSiON,AddingMemory}, as well as a high photoluminescence quantum efficiency and good thermal stability \cite{ColorPointMSiON}, enabling a high TL signal strength at elevated temperatures. 
The method is extendable to other materials when the trap depth distribution can be translated into TL.  Accurate determination of the trap depth distribution will lead to a step forward in understanding the properties of persistent and storage phosphors.

\section{Materials and methods}\label{sec:matermethod}
The BaSi$_2$O$_2$N$_2$:2\%{Eu}$^{2+}$ phosphor was prepared by a two-step solid-state reaction method \cite{step}, according to
\begin{linenomath*}
\begin{subequations}
\begin{align}
& 1.96\text{BaCO}_3 + \text{SiO}_2  + 0.02\text{Eu}_2\text{O}_3 \nonumber \\ 
&\quad \qquad \qquad \rightarrow \text{Ba}_2\text{SiO}_4\text{:2\%}\text{Eu}^{2+}  + 1.96\text{CO}_2, \label{eqn:BaSiO}  \\
& \text{Ba}_2\text{SiO}_4\text{:2\%}\text{Eu}^{2+} + \text{Si}_3\text{N}_4 \rightarrow 2\text{BaSi}_2\text{O}_2\text{N}_2\text{:2\%}\text{Eu}^{2+}. \label{eqn:BaSiON} 
\end{align}
\end{subequations}
\end{linenomath*}
The raw materials BaCO$_3$ (99.8 \%, 1 \textmu m powder, Alfa Aesar), SiO$_2$ (99.5\%, 325 mesh powder, Alfa Aesar), and Eu$_2$O$_3$ (99.9\%, Alfa Aesar) were used in stoichiometric amounts except that 103\% Si$_3$N$_4$ ($\alpha$ phase, 99.9\%, 1 \textmu m powder, Alfa Aesar) was supplied to facilitate the reduction of Eu$^{3+}$ to Eu$^{2+}$ \cite{SiN_reducingAgent}. The sintering temperature and duration for Eqs. \ref{eqn:BaSiO} and  \ref{eqn:BaSiON} were 1200 {\textdegree}C, 4 h and 1450 {\textdegree}C, 10 h, respectively. A 94\% N$_2$-6\% {H}$_2$ forming gas atmosphere was applied at a constant rate (0.16 L min$^{-1}$) during the entire thermal process. The product was crushed and ground to fine powders, and then washed by diluted hydrogen chloride (HCl,$<$1 vol\%). After being dried at 80 {\textdegree}C for at least 10 h, BaSi$_2$O$_2$N$_2$:2\%{Eu}$^{2+}$ powders were ready for further use.

A TQ profile was collected to correct TL curves by using the method proposed in Ref. \cite{Anorthite}. The spectra were acquired by a home-built setup \cite{jonasb_trapping_detrapping_SAO}. The excitation light of 370 nm [full-width-half-maximum (fwhm) 5 nm] was from a Xe arc lamp equipped with a monochromator, while the emission was collected by an EMCCD camera (Princeton Instruments ProEM 1600) coupled to a spectrograph (Princeton Instruments Acton SP2300). The integration time was 1 s. The phosphor was cooled to 213 K and then heated to 498 K at a step of 5 K, with optical excitation at each temperature $T$ for 30 s. For each $T$, five spectra from the time range from 24 to 28 s were averaged  to represent the PL emission intensity $I(T)$ (Supplemental material (SM) \cite{SM}, Sec. I). For each TL curve, the measured TQ was linearly interpolated at each temperature recording of the TL curve. 

\begin{figure}
\centering
\includegraphics[width = 0.84\linewidth]{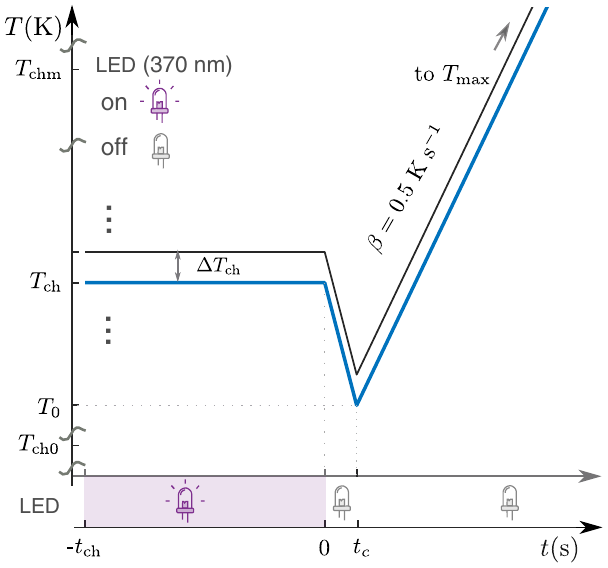} 
\caption{\textbf{The procedure for a TL experiment}. The phosphor was charged by UV light at variable charging temperature $T_\text{ch}$ with a fixed charging duration $t_\text{ch}$. After lowering the temperature to $T_0\leq T_\text{ch}-30 \text{K}$, the phosphor will be heated to temperature $T_\text{max}$ at a fixed heating rate $\beta$ (0.5 K s$^{-1}$). Here, $T_\text{ch}=[T_\text{ch0}:\Delta T_\text{ch}: T_\text{chm}]$, with  $\Delta T_\text{ch}=5\text{ K},T_\text{ch0}=223\text{ K}, \text{ and } T_\text{chm} = 393\text{ K}$.} 
\label{fig_chargingprotocol} 
\end{figure}

The experimental procedure that allows us to extract trap depth distributions uses fixed irradiance, fixed charging duration, and variable charging temperature $T_\text{ch}$ (Fig. \ref{fig_chargingprotocol}). The ultraviolet (UV) excitation light (370 nm, FWHM 20 nm) was from a light-emitting-diode (LED) which was driven by a current of 50 mA. The charging temperature was set in the range $T_\text{ch}=[T_\text{ch0}:\Delta T_\text{ch}: T_\text{chm}]$, with  $\Delta T_\text{ch}=5\text{ K},T_\text{ch0}=223\text{ K}, \text{ and } T_\text{chm} = 393\text{ K}$. To acquire a TL curve, the phosphor was first excited at $T_\text{ch}$ for 300 s and then cooled down at a rate of  $-0.5$ K s$^{-1}$ to $T_0$ ($T_0\leq T_\text{ch}-30 \text{ K}$), where the TL intensity is negligible. Finally, the phosphor was heated up to 493 K at a heating rate $\beta=$ 0.5 K s$^{-1}$. The emission was detected by a photometer (International Light Technologies, ILT1700) equipped with a photopic filter (YPM). Every TL curve was corrected by the TQ profile to account for the nonradiative decay of the luminescent activators as well. 

\section{Results}
\begin{figure}
\centering
\includegraphics[width = 0.98\linewidth]{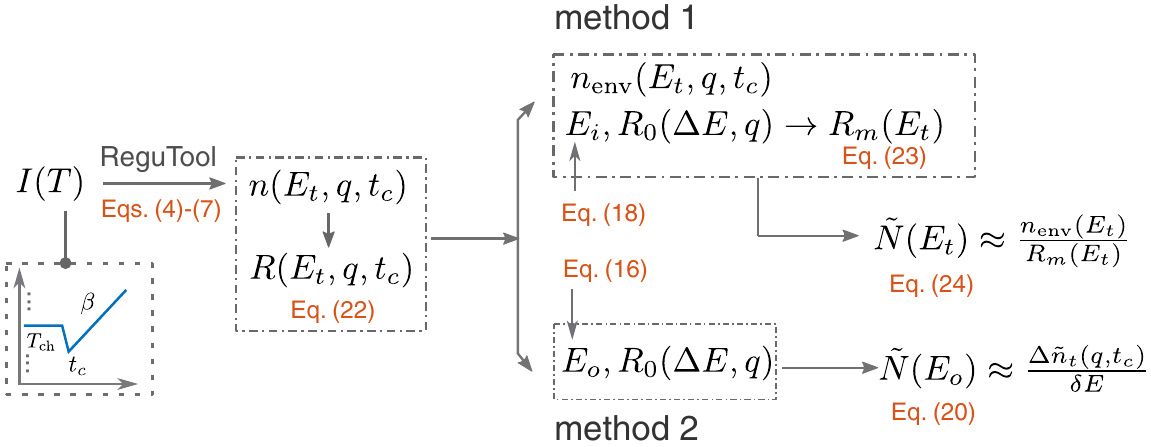} 
\caption{\textbf{Flowchart of the method}. The TL curves are first converted into electron population functions, and the relative filling functions are approximated. Two methods can be chosen to calculate the trap depth distribution $N(E_t)$.} 
\label{fig_flowchart} 
\end{figure}

We briefly outline the procedure of extracting a trap depth distribution (see Fig. \ref{fig_flowchart}), which is the focus of the following sections. In Sec. \ref{sec:n}, the electron population functions $n(E_t,q,t_c)$  are reconstructed from the experimental TL curves from a carefully designed charging procedure (see Sec. \ref{sec:matermethod}). A numerical recipe called the Tikhonov regularization method is used to solve this inverse problem within the framework of first-order kinetics of TL (see Eqs. \ref{eqns_TLregu}-\ref{eqn_Vnhat}). The presence of a thermal barrier $\Delta E$ can be inferred from these electron population functions. In Sec. \ref{sec:f}, the filling function (Eq. \ref{eqn_Rdef}) is calculated, using first-order kinetics for the trapping and recombination processes during the charging process. From simulations of filling functions, a method is proposed to approximate the magnitude of the filling function $f_0(\Delta E,q)$. The subtle relationship between $f_0(\Delta E,q)$ and $n(E_t,q,t)$ on the one hand and the trap depth distribution $N(E_t)$ on the other hand can be revealed accordingly. In Sec. \ref{sec:fexp}, the trap depth distribution of BaSi$_2$O$_2$N$_2$:2\%{Eu}$^{2+}$ is then finally reconstructed. Two equivalent methods are demonstrated, reaching consistent results.
 
\subsection{Electron population function}\label{sec:n}
Electron population functions after charging are obtained from TL glow curves. First-order kinetics is assumed for TL, i.e., retrapping of electrons among traps is assumed to be negligible. This is a valid assumption during a TL experiment because the rate coefficient of detrapping increases substantially with increasing temperature. (See Sec. \ref{sec: localmodel} for details.) Assuming first-order kinetics thus leads to a convenient means to infer the information of a phosphor during/after charging.

\subsubsection{Tikhonov regularization method}\label{sec:kernel}
According to first order kinetics, the TL intensity from an electron population function $n(E_t,q,t_c)$ is given by the Fredholm integral of the first kind \cite{fredholm-1903,ChenMcKeever97book},
\begin{linenomath*}
\begin{subequations}
\begin{align}
I(T) &= C\times\int_0^{\infty}n(E_t,q,t_c)K(E_t,T)dE_t, \label{eqn_TLFET}\\
K(E_t,T) &=\frac{\nu_r}{\beta}\textrm{exp}\left[-\frac{E_t}{k_{\textsc{b}}T}-F(E_t,T)+F(E_t,T_0)\right], \label{eqn_kernelEtT} \\
F(E_t,T) &= \frac{\nu_r}{\beta}\int_{0}^{T}\textrm{exp}\left(-\frac{E_t}{k_\textsc{b}T'}\right)dT', \label{eqn_Tintegral}
\end{align}
\label{eqns_TLregu}
\end{subequations}
\end{linenomath*}
where the function $K(E_t,T)$ is referred to as the kernel that translates the electron population function $n(E_t,q,t_c)$ to the TL intensity $I(T)$ and $F(E_t,T)$ is often called the \textit{temperature integral} \cite{Rev_TintegralFlynn}. Here, $C$ is a coefficient to render the appropriate unit for $I(T)$. The meaning of $t_c, T_0,\text{ and }\beta$ have been elucidated in section \ref{sec:matermethod} (see Fig. \ref{fig_chargingprotocol}). Equation \ref{eqn_TLFET} was proposed  by Randall \textit{et al.} \cite{RandallWilkins_I} and  Randall and Willkins \cite{RandallWilkins_II}, but the electron population function $n(E_t,q,t_c)$ was replaced by a trap depth distribution $N(E_t)$. \par

An analytic expression for $F(E_t,T)$ was proposed by M. Balarin \cite{TempInt_Balarin},
\begin{equation}
F(E_t,T) = \frac{\nu_r}{\beta}\frac{k_{\textsc{b}}T^2}{E_t}\textrm{exp}\left(-\frac{E_t}{k_{\textsc{b}}T}\right)\frac{1}{\sqrt{1 + 4k_{\textsc{b}}T/E_t}}, \label{eqn_FBalarin}
\end{equation}
which offers high accuracy even when $E_t/k_\textsc{b}T$ is small \cite{Rev_TintegralOrfao}.This formula provides an easy algorithm for numerical evaluation of the temperature integral (Eq. \ref{eqn_Tintegral}) in computing the kernel.

\begin{figure}
\centering
\includegraphics[width = 0.98\linewidth]{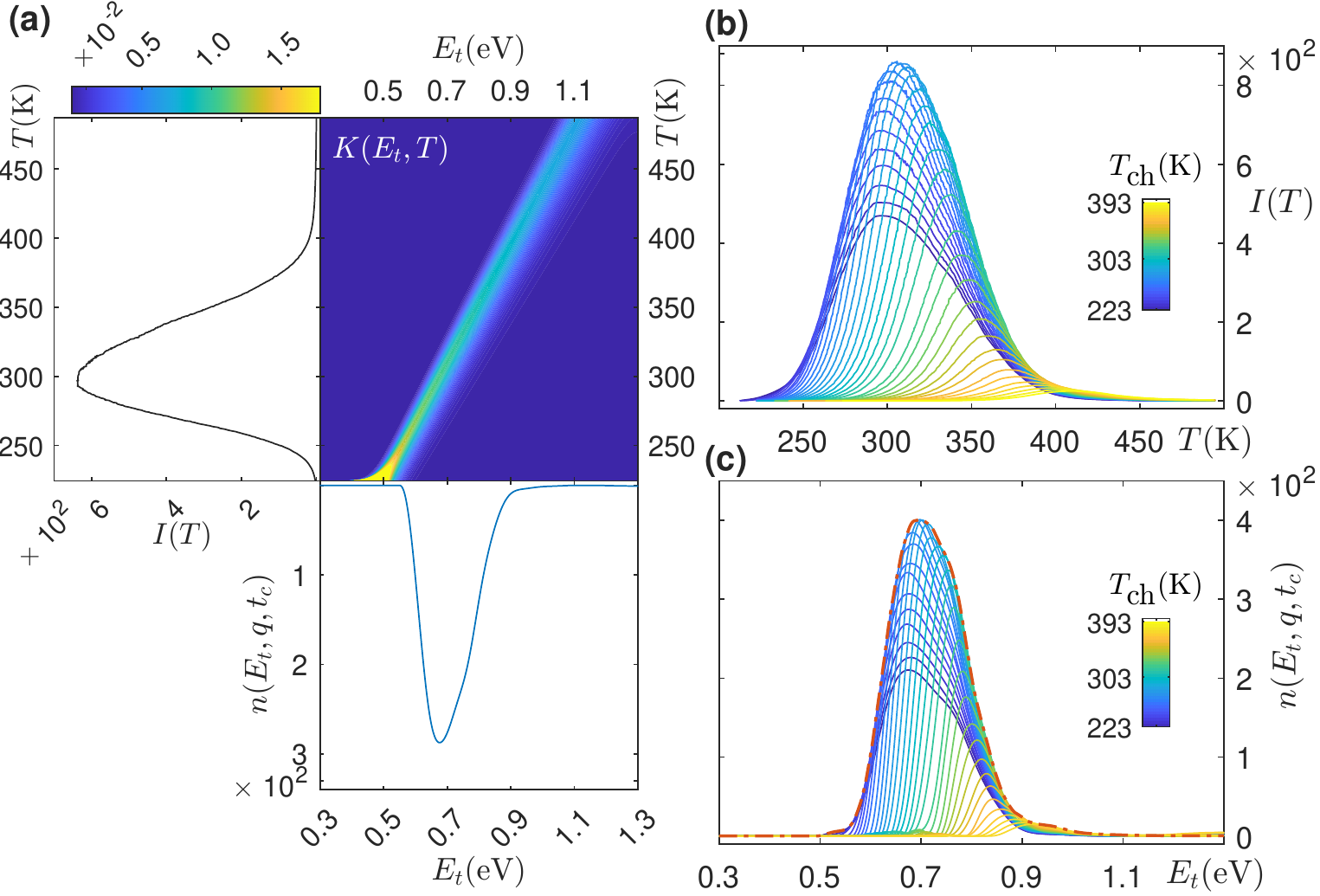} 
\caption{\textbf{The electron population function $n(E_t,q,t_c)$}. \textbf{(a)} The kernel $K(E_t,T)$ maps the TL curve $I(T)$ into the electron population function $n(E_t,q,t_c)$ via discrete regularization method Eqs. \ref{eqns_TLregu}-\ref{eqn_Tintegral} ($T_\text{ch}=243\text{ K}$). For variable charging temperature $T_\text{ch}$, \textbf{(b)} the TL curves can thus turn into \textbf{(c)} the electron population functions $n(E_t,q,t_c)$, from which an envelope can be constructed accordingly (the orange line). Note $\nu_r=10^{10}$ $\text{s}^{-1}$.} 
\label{fig_TLepop} 
\end{figure}
 
The electron population function $n(E_t,q,t_c)$ can be numerically obtained by solving Eqs. \ref{eqns_TLregu}-\ref{eqn_Tintegral} with the formula Eq. \ref{eqn_FBalarin}. As a first step, the temperature and energies are  discretized over a grid $[T_0, T_m]\times[E_a, E_b]$, leading to
\begin{equation}
\label{eqn_Kni}
Kn = I, 
\end{equation}
which is equivalent to Eq.~\ref{eqn_TLFET}. Herein, $K$, $I$ and $n$ are matrices representing the kernel, experimental TL data and electron population function, respectively. The integral equation Eq. \ref{eqns_TLregu} and its discrete counterpart Eq. \ref{eqn_Kni} are ill-conditioned, implying that approximate solutions are possible if the Picard condition is fulfilled \cite{HansenPicardcond}.  Therefore, no stable solution for $n$ can be extracted from Eq. \ref{eqn_Kni} by linear least squares methods, i.e., seeking $\hat{n}$ that minimizes the residual norm squared $\lVert K\hat{n}-I \lVert_2^2$. The Tikhonov regularization method can solve this inverse problem by minimizing the functional \cite{tikhonovBook77,Hansen_reguBook}, 
\begin{equation}
V(\hat{n})=\lVert K\hat{n} - I\rVert _2^2+ {\lambda}^2 \lVert L\hat{n}\rVert_2^2
\label{eqn_Vnhat}
\end{equation}
in which $\lambda$ is the regularization parameter, and  $L$ is the discrete approximation of a derivative operator. Additionally, a non-negativity constraint is imposed for the solution, i.e., $\hat{n}\geq 0$. This regularization operator favors smooth solutions for $\hat{n}$ (small derivatives), leading to an improved numerical stability of the solution. The smoothness of $\hat{n}$ is hence implicitly assumed during Tikhonov regularization, and agrees with the physical picture of electron population functions. The value of $\lambda$ fixes the relative weights of both contributions in the minimization and is numerically chosen to guarantee a good balance between regularization and agreement to experiment \cite{L_curveHasen}. In this work, the Tikhonov regularization is implemented via the \textsc{regularization tools matlab} package \cite{RegToolsRef0,RegToolsRef}. More details are given in Sec. II of SM \cite{SM}.  \par

The kernel $K(E_t,T)$ maps TL curves into electron population functions via the Tikhonov regularization method. An individual example is shown in Fig. \ref{fig_TLepop}a, for which $T_\text{ch}=243\text{ K}$. The experimental TL curves and the extracted electron population functions are displayed in Fig. \ref{fig_TLepop}b and Fig. \ref{fig_TLepop}c for $T_\text{ch}=[T_\text{ch0}:\Delta T_\text{ch}: T_\text{chm}]$, with  $\Delta T_\text{ch}=5\text{ K},T_\text{ch0}=223\text{ K}, \text{ and } T_\text{chm} = 393\text{ K}$. Obviously, the higher $T_\text{ch}$, the further the tails of electron population functions extend, suggesting a temperature dependent filling of traps. Furthermore, an envelope of the electron population functions, i.e. $n_\text{env}(E_t,q,t_c)$, can be calculated by an interpolation method (Sec. III of SM \cite{SM}). Shown as the orange line in Fig. \ref{fig_TLepop}c, the envelope $n_\text{env}(E_t,q,t_c)$ will be crucial to reconstruct high-precision trap depth distribution, which will be discussed in Sec. \ref{sec:f}.

\subsubsection{The presence of a thermal barrier} \label{sec:prescencedE}
The presence of a thermal barrier for charging can be revealed qualitatively. For each electron population function $n(E_t,q,t_c)$, the total number of trapped electrons per unit volume can be calculated by
\begin{equation}
n_t(q,t_c)=\int_{E_a}^{E_b}n(E_t,q,t_c)dE_t,\label{eqn_nt}
\end{equation}
and the corresponding difference $\Delta n_t(q,t_c)$ can be calculated as
\begin{equation}
\Delta n_t(q,t_c) = n_t(q_1,t_c)-n_t(q,t_c).\label{eqn_dnt}
\end{equation}
Herein, $q$ can equally be replaced by $T_\text{ch}$, i.e. $q\rightarrow T_\text{ch}$ and $q_1 \rightarrow T_\text{ch}-\Delta T_\text{ch}$, since $I_e$ and $t_\text{ch}$ are already fixed. \par

\begin{figure}
\centering
\includegraphics[width = 0.98\linewidth]{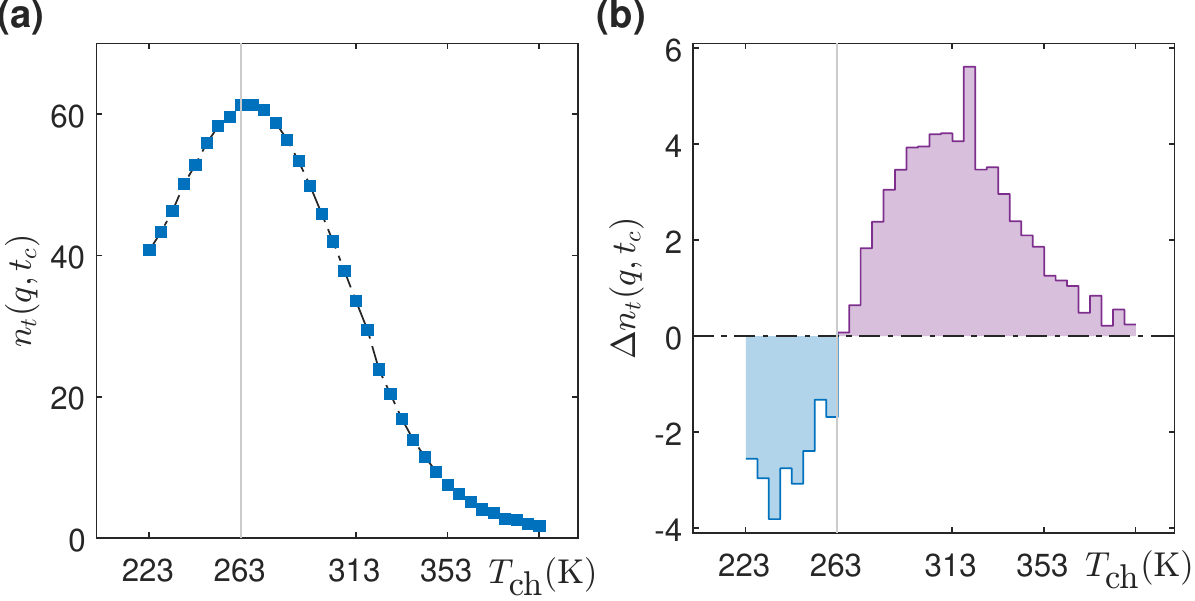} 
\caption{\textbf{The existence of $\Delta E$}. \textbf{(a)} The total number of trapped electrons per unit volume $n_t(q,t_c)$ as a function of $T_\text{ch}$ reaches a peak at $T_\text{ch}\approx 263 \text{ K}$. \textbf{(b)} As $T_\text{ch}$ decreases, the corresponding difference $\Delta n_t(q,t_c)$ turns from positive into negative at $T_\text{ch}\approx 263 \text{ K}$.}
\label{fig_DeltaE} 
\end{figure}

When $\Delta E = 0.0 \text{ eV}$ (i.e., in the absence of a thermal barrier for charging), $n_t(q,t_c)$ should be a non-decreasing function with decreasing $T_\text{ch}$ because more shallow traps can be filled at lower temperature. To put it another way, the corresponding difference $\Delta n_t(q,t_c)$ will always be non-negative. This is not the case for the phosphor under study, as shown in Fig. \ref{fig_DeltaE}. At $T_\text{ch}\approx 263\text{ K},\Delta n_t(q,t_c)$ turns from positive to negative, which means the traps are already less efficiently filled as $T_\text{ch}$ decreases. This suggests the presence of a thermal barrier for charging.

\subsection{Filling function}\label{sec:f}
The kinetics of electronic transitions during charging is required to reveal information on the filling functions after charging. This information is needed to find a recipe to calculate filling functions from the electron population functions. In this section, the trapping and recombination processes are assumed to take place within isolated pairs and thus first-order kinetics is hypothesized naturally (Sec. \ref{sec: localmodel}). Given appropriate parameters, the filling function can be simulated under the proposed charging procedure. The method for extracting the filling functions from the electron population functions is then revealed from the simulation, and two methods of reconstructing trap depth distributions are proposed accordingly (Sec. \ref{sec:analysisfilling}).

\subsubsection{Analytic expression} \label{sec: localmodel}
\begin{figure}
\centering
\includegraphics[width = 0.92\linewidth]{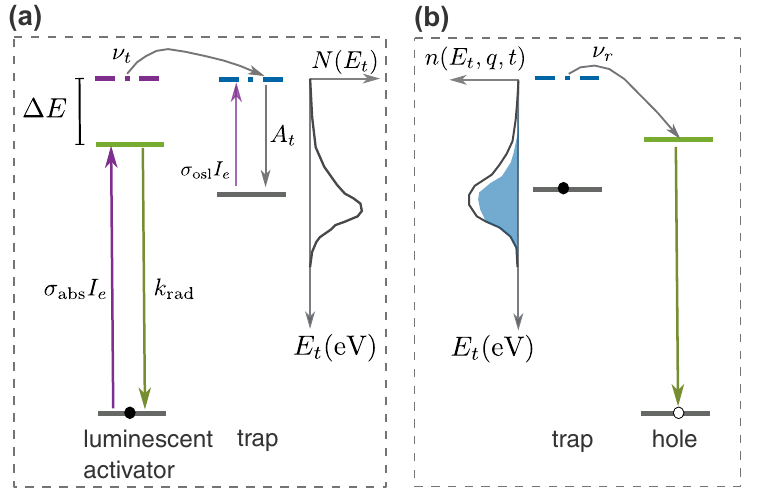} 
\caption{\textbf{Local model for trapping and recombination}. The isolated-pair approximation is assumed for trapping and recombination. \textbf{(a)} The trapping process takes place between an isolated luminescent activator and an empty trap, i.e., an activator-trap pair. \textbf{(b)} Recombination takes place between an isolated filled trap and a hole, i.e., an electron-hole pair. Parameters are displayed for important electron transitions. The trap depth distribution $N(E_t)$ and electron population function $n(E_t,q,t_c)$ are shown as gray curve and blue filled area, respectively.} 
\label{fig_trapcenter} 
\end{figure}

It is typically reckoned that two different species are involved in persistent luminescence and thermoluminescence processes: the luminescent activators and the traps. A luminescent activator, e.g., Eu$^{2+}$, has a ground state and a dense manifold of excited states \cite{joos-2020-1}. In the kinetic models for TL, electronic states are typically regarded in a mean-field single-electron approximation, leading to few orbitals that a charge carrier can occupy or not \cite{ChenMcKeever97book}. As such, a luminescent activator is usually approximated by one ground state and one excited state, leading to a four-orbital energy level scheme for which equilibrium occupations can be modeled via the Fermi-Dirac distribution. An electron in the excited orbital of the luminescent activator either decays radiatively to the ground state of the luminescent activator or gets captured at a trap if it is able to overcome the thermal barrier $\Delta E$ (Fig. \ref{fig_trapcenter}a). Chemically, a trap can be a lattice defect, e.g. an oxygen vacancy \cite{CaAlO_EuR} or even a co-dopant, like in the case of Dy in Sr$_4$Al$_{14}$O$_{25}$: Eu$^{2+}$,Dy$^{3+}$ \cite{DyasTrap}. If an electron trap is empty, it can capture an electron. If the trap is  filled, it can supply an electron to recombine with a hole nearby, provided that the electron can overcome the thermal barrier $E_t$, i.e. the trap depth (Fig. \ref{fig_trapcenter}b). This detrapping process is referred to as recombination. The hole in persistent phosphors is often reckoned as immobile since its mobility is much smaller than that of electrons. In the case of Eu$^{2+}$-based persistent phosphors, the hole is localized at the (photo-)oxidized Eu$^{2+}$, i.e. the Eu$^{3+}$ center. In the remainder, the situation where an electron in a filled trap is transferred to recombine with a (localized) hole is referred to as an \textit{electron-hole pair}. Furthermore, the process where an electron is transferred from a filled trap to another empty trap, i.e., retrapping, is not considered.

Assuming first-order kinetics in a local model is very reasonable. Firstly, the density of  luminescent activators (usually of the order of 1 mol\%) is often greater than that of empty traps when they are crystallographic defects, making that activator-trap pairs are on average far apart. Secondly, multi-electron \textit{ab initio} calculations have shown that electron transfer between activators and traps occurs locally, not requiring the delocalization of a charge carrier to a conduction band state \cite{joos-2021-1, joos-2021-2}. Thirdly, empty traps can only be filled to a limited level by exciting the luminescent activators, even if the density of traps is high, as is the case of Dy in Sr$_4$Al$_{14}$O$_{25}$:Eu$^{2+}$,Dy$^{3+}$ \cite{DyasTrap}, making that electron-hole pairs are thus far apart. First order kinetics hence emerges naturally since the involved pairs are to a large extent isolated and independent. An isolated activator-trap pair can only transform into an electron-hole pair upon trapping, while the electron-hole pair turns into an activator and an empty trap upon recombination. Here, it is assumed that the charge transfer dominantly takes place within pairs with the shortest separation. Under this assumption, the density of electron-hole pairs is proportional to the density of filled traps \cite{TunnelingKClPRB}. Therefore, the filling function $f(E_t,\Delta E,q,t)$ itself follows the first-order ordinary differential equation, 
\begin{linenomath*}
\begin{align}
\frac{\partial f(E_t,\Delta E,q,t)}{\partial t} &= k_\text{trap}(\Delta E, q)\left[1-f(E_t,\Delta E,q,t)\right] \nonumber \\
&\quad-k_\text{rcb}(E_t,q)f(E_t,\Delta E,q,t), \label{eqn_fillODE} 
\end{align}
\end{linenomath*}
where $k_\text{trap}(\Delta E, q)$ and $k_\text{rcb}(E_t,q)$ are the trapping coefficient and recombination coefficient, respectively.

The trapping and recombination coefficients in Eq. \ref{eqn_fillODE} can in principle be deduced by analyzing the kinetics of elementary trapping and recombination events. At the time scale needed to register a single TL data point (often $>10$ ms), the density of electrons in the excited state of a luminescent activator or a trap will have reached its maximum change upon any abrupt perturbation. The relative values of the trapping and recombination coefficients can thus analyzed instead (depicted in Fig. \ref{fig_trapcenter}). According to the analysis in Sec. IV of SM \cite{SM}, the trapping and recombination coefficients read, 
\begin{linenomath*}
\begin{align}
k_\text{trap}(\Delta E,q) &= \nu_t \textrm{exp}\left(-\frac{\Delta E}{k_\textsc{b}T}\right)\frac{\sigma_\text{abs}I_e(\lambda)}{k_\text{rad}},\label{eqn_ktrap} \\
k_\text{rcb}(E_t,q) &= \frac{A_t}{A_t+\nu_r}\left[\nu_r \textrm{exp}\left(-\frac{E_t}{k_\textsc{b}T}\right) +\frac{\nu_r}{A_t}\sigma_\text{osl}I_e(\lambda)\right], \label{eqn_krecomb}
\end{align}
\end{linenomath*}
respectively. Here, $\sigma_\text{abs}$ is the optical absorption cross-section of Eu$^{2+}$, and $k_\text{rad}$ is the spontaneous emission coefficient of the emitting $4f^65d^1$ state of Eu$^{2+}$. Similarly for traps, $\sigma_\text{osl}$ is the absorption cross section of optically stimulated detrapping, while $A_t$ is the de-excitation coefficient of the excited state of the trap, regardless of its depth. The frequency factors $\nu_r$ and $\nu_t$ correspond to the trapping and recombination processes, respectively.  

For a phosphor with all traps  initially empty, an optical charging under fixed irradiance $I_e(\lambda)$, after duration $t_\text{ch}$ and at temperature $T_\text{ch}$ leads to the filling function as a solution of Eq. \ref{eqn_fillODE},
\begin{linenomath*}
\begin{align} 
\label{eqn_fillt0}
& f(E_t,\Delta E, q, t=0)  = \frac{k_\text{trap}(\Delta E, q)}{k_\text{trap}(\Delta E, q)+k_\text{rcb}(E_t, q)} \nonumber \\
&\qquad \times\left\{1-\textrm{exp}\left[-\left(k_\text{trap}(\Delta E, q) + k_\text{rcb}(E_t, q)\right) t_\text{ch}\right]\right\}.
\end{align}
\end{linenomath*}
Clearly, the magnitude and shape of the filling function are influenced by the thermal barrier $\Delta E$.  The cooling to $T_0$ after charging (see Fig. \ref{fig_chargingprotocol}) further reduces the filling function to 
\begin{linenomath*}
\begin{align}
f(E_t,\Delta E, q,t_c) &= f(E_t,\Delta E,q,t=0) \nonumber \\
&\quad \times\textrm{exp}\left[-F(E_t,T_0)+F(E_t,T_\text{ch})\right].\label{eqn_fillcool}
\end{align}
\end{linenomath*}

It is unrealistic to fit this model directly to experimental observations because there is a huge number of parameters, many of which are not easily available. Instead, parameters are provided from experiments or estimated, as shown in Table \ref{tab:parametersimulation}. For example, the thermal barrier for charging $\Delta E$ takes an arbitrary value of 0.255 eV. Actually, the exact value of $\Delta E$ is not important in extracting the trap depth distribution because it dominantly influences the magnitudes of filling functions, which will cancel out (see Sec. \ref{sec:uncertaintyAnalysis}). Simulations of Eqs. \ref{eqn_fillt0} and \ref{eqn_fillcool} are conducted to analyze the filling functions under the designed charging procedure in Sec. \ref{sec:analysisfilling}. From this analysis, methods of extracting trap depth distribution will be proposed subsequently in Sec. \ref{sec:2methods}.

\begin{table}
\caption{\label{tab:parametersimulation}The parameters for simulations}
\begin{ruledtabular}
\begin{tabular}{lllll}
Parameter & Unit & Value &Comment\\
\midrule
$\Delta E$  & eV & 0.255 &  \\ 
$\sigma_\text{abs}$ & cm$^{2}$ &$3\times10^{-18}$ &  \\
$\sigma_\text{osl}$  &cm$^{2}$ &$10^{-17}$ &  \\
$k_\text{rad}$ &s$^{-1}$ &1.54$\times 10^6$ & see Ref. \citep{ColorPointMSiON}  \\
$\nu_r$ & s$^{-1}$ &$10^{10}$ &  \\
$\nu_t$ & s$^{-1}$ &$10^{10}$ &  \\ 
$A_t$ & s$^{-1}$ &$10^{12}$ & \\
$I_e(\lambda)$ & $\frac{\text{photons}}{\text{cm}^{2}\text{s}}$ & 5$\times 10^{15}$ & $\lambda = 370$ nm \\
$T_\text{ch}$ & K &-  & \footnotemark[1] \\
$t_\text{ch}$ & s &-  & \footnotemark[2] \\
$k_{\textsc{b}}$ & eV K$^{-1}$ &8.617$\times10^{-5}$  &  \\
\end{tabular}
\end{ruledtabular}
\footnotetext[1]{The values are specified in the figures or their captions.}
\footnotetext[2]{The values are specified in the figures or their captions.}
\end{table}

\subsubsection{Analysis of filling functions}\label{sec:analysisfilling}
By using parameters in Table \ref{tab:parametersimulation}, the filling function $f(E_t,\Delta E,q,t_c)$ (Eq. \ref{eqn_fillcool}) was simulated at variable charging temperature $T_\text{ch}$, with fixed charging duration $t_\text{ch}$, fixed charging irradiance $I_e$, and $T_0=T_\text{ch}-40\text{ K}$ (cooling rate $-1.0$ K s$^{-1}$). The filling functions for $\Delta E=0.0\text{ eV and } \Delta E = 0.255 \text{ eV}$ are discussed in the following to reveal the methods of extracting trap depth distributions.

\begin{figure}
\centering
\includegraphics[width = 0.98\linewidth]{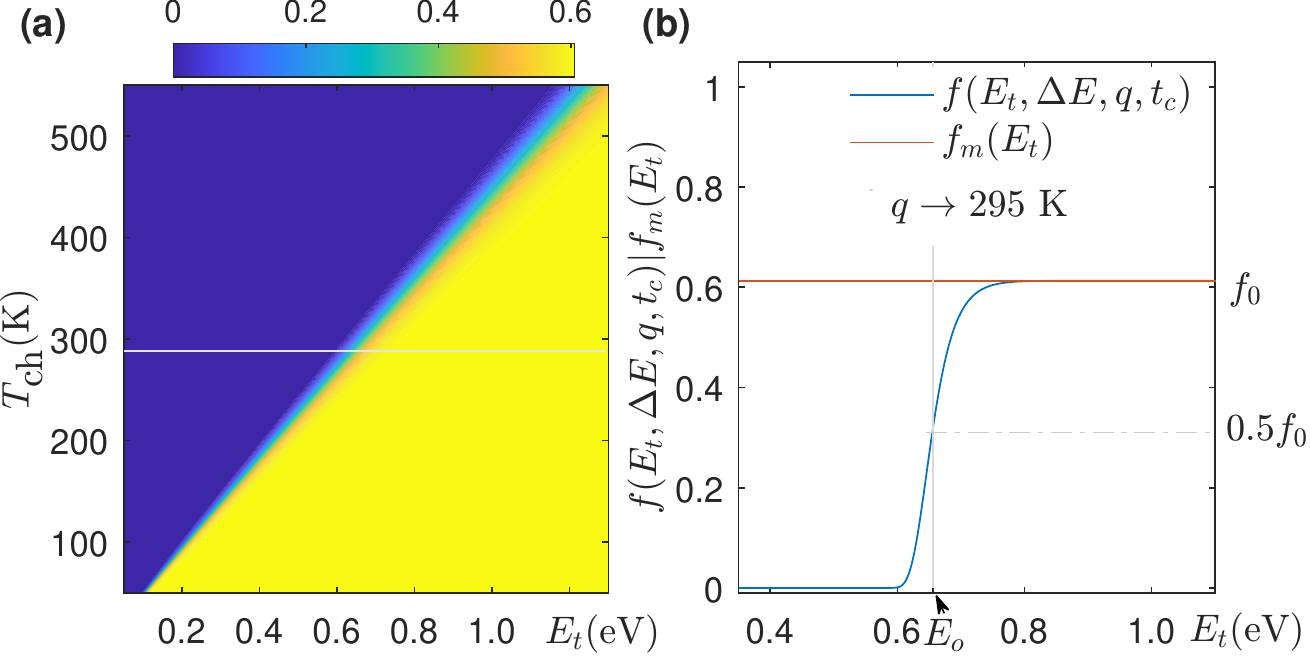} 
\caption{\textbf{The filling function in the case without a thermal barrier for charging ($\Delta E = 0.0$ eV)}. \textbf{(a)} The filling function $f(E_t,\Delta E, q,t_c)$ in the $T_\text{ch}\times E_t$ plane ($q\rightarrow T_\text{ch}$) suggests the magnitude function $f_m(E_t)$ is independent of charging temperature $T_\text{ch}$. The gray line indicates the filling function for $q\rightarrow295\text{ K}$. \textbf{(b)} Indeed, the magnitude function $f_m(E_t)$ is independent of charging temperature $T_\text{ch}$ (orange line). An individual filling function (e.g., the one for $q\rightarrow295$ K) characterizes a magnitude of filling $f_0$ and a characteristic trap depth $E_o$. Note that $t_\text{ch}=0.01\text{ s}$.} 
\label{fig_fillFuncNOdE} 
\end{figure} 

\textbf{a. $\Delta E = 0.0 \text{ eV}$}. In this case, the magnitudes of the filling functions are independent of their corresponding charging temperature $T_\text{ch}$. The color plot of $f(E_t,\Delta E,q,t_{c})$ in the $T_\text{ch}\times E_t$ plane is shown in Fig. \ref{fig_fillFuncNOdE}a, where the filling function for $q\rightarrow \text{ 295 K}$ is indicated by the gray line. As shown in Fig. \ref{fig_fillFuncNOdE}b, an individual filling function can be reduced to its magnitude $f_0$ and a characteristic trap depth $E_o$. As such, it can be approximated by a Heaviside step function, i.e.,
\begin{linenomath*}
\begin{equation}
f(E_t,\Delta E,q,t_{c})\approx f_0(\Delta E,q)H[E_t-E_o(q)].
\label{eqn_NOdEHeavyside}
\end{equation}
\end{linenomath*}
Consequently, the characteristic trap depth $E_o$ can be determined by
\begin{linenomath*}
\begin{equation}
 f(E_o,\Delta E,q,t_c)= 0.5 f_0(\Delta E,q). \label{eqn_calEo}
\end{equation}
\end{linenomath*}
Therefore, the pair of $E_o$ and $f_0(\Delta E,q)$ represents the corresponding filling function under the charging condition $q$. When (the magnitudes of) all filling functions for different charging temperatures are collected, a function $f_m(E_t)$ can be constructed that gives the magnitude of the filling functions as a function of variable charging temperature ($q\rightarrow T_\text{ch}$),
\begin{linenomath*}
\begin{equation}
f_m(E_t):E_t(q)\mapsto f_0(\Delta E,q).
\label{eqn_fmEtdef}
\end{equation} 
\end{linenomath*}
Here, $f_m(E_t)$ is called the \textit{magnitude function}. From Fig. \ref{fig_fillFuncNOdE}b, the magnitude function $f_m(E_t)$ (orange line) is independent of $E_t$ and thus of charging temperature in the case at hand.

A method of extracting trap depth distributions becomes visible in this case. Given the charging condition $q\rightarrow T_\text{ch}$ and $q_1\rightarrow T_\text{ch}-\Delta T_\text{ch}$, the difference of the total number of traps per unit volume $\Delta n_t(q,t_c)$ (Eq. \ref{eqn_dnt}) is proportional to $N(E_t)$ for $E_t\in[E_t(q_1),E_t(q)]$. No correction is needed for either the magnitude $f_m(E_t)$ or the electron population function $n(E_t,\Delta E,q,t_{c})$. This is the essence of the method from Ref. \cite{Koen_trap_depth}, in which $E_o$ is extracted by the initial rise method. The current method outperforms the one in Ref. \cite{Koen_trap_depth} because of its higher precision in extracting $E_o$.

\begin{figure}
\centering
\includegraphics[width = 0.98\linewidth]{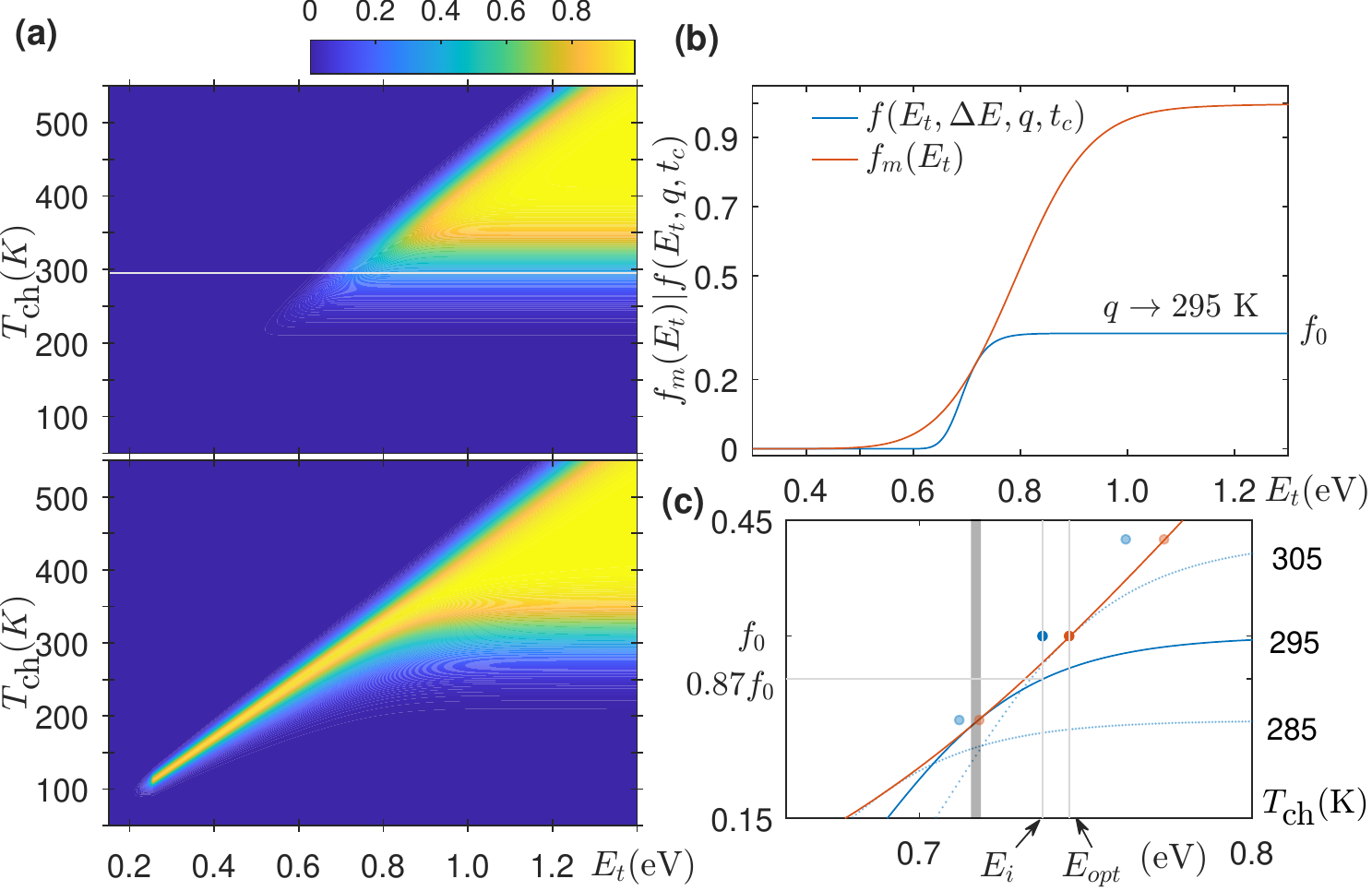} 
\caption{\textbf{The filling function in the case of a thermal barrier for charging ($\Delta E=0.255\text{ eV}$)}. \textbf{(a)} The  filling function $f(E_t,\Delta E,q,t_c)$ (top panel) and the normalized filling function $f(E_t,\Delta E,q,t_c)/f_m(E_t)$ (bottom panel) in the $T_\text{ch}\times E_t$ plane ($q\rightarrow T_\text{ch}$) suggest an optimal range of charging temperature for a given trap depth. \textbf{(b)} The magnitude function $f_m(E_t)$ is tangent to the filling function $f(E_t,\Delta E,q,t_c)$. \textbf{(c)} The magnitude function $f_m(E_t)$ can be approximated by the characteristic trap depth $E_i$ and the magnitude $f_0(\Delta E,q)$ of the filling function. Note $t_\text{ch} = 100$ s.} 
\label{fig_fillFuncdE} 
\end{figure}

\textbf{b. Non-zero $\Delta E$}. The filling function with non-zero $\Delta E$ clearly indicates the dependence of the magnitude function $f_m(E_t)$ on the charging temperature $T_\text{ch}$. The color plot for $f(E_t,\Delta E,q,t_c)$ with $\Delta E = 0.255 \text{ eV}$ (Fig. \ref{fig_fillFuncdE}a, top panel) shows that the filling of traps depends on the charging temperature $T_\text{ch}$. Every filling function $f(E_t,\Delta E,q,t_c)$ can be normalized by its magnitude $f_0(\Delta E,q)$, and the color plot of the normalized filling function is presented in Fig. \ref{fig_fillFuncdE}a  as $f(E_t,\Delta E,q,t_c)/f_m(E_t)$ (bottom panel). For a given trap depth, there exists an optimal range of charging temperature  that optimizes the output of persistent luminescence, which has been observed in many phosphors \cite{jrdu_Topt}. The magnitude function $f_m(E_t)$ (Fig. \ref{fig_fillFuncdE}b), which is tangent to the filling functions, increases with increasing trap depth, indicating the presence of a thermal barrier for charging $\Delta E$.   \par

The magnitude function $f_m(E_t)$ carries two important implications, which are crucial for the extraction of trap depth distributions. 

The first implication is that it can be approximated from filling functions, which is illustrated in Fig. \ref{fig_fillFuncdE}c. For a given filling function $f(E_t,\Delta E,q,t_c)$, the magnitude $f_0(\Delta E,q)$ can be easily extracted and it corresponds to a theoretical trap depth $E_{opt}$ which satisfies,
\begin{linenomath*}
\begin{equation*}
f_m(E_{opt}) = f_0(\Delta E,q). \label{eqn_Eopt}
\end{equation*}
\end{linenomath*}
However, $E_{opt}$ is inaccessible experimentally because the magnitude function is unknown beforehand. By using the filling function alone, a characteristic trap depth $E_i$ is used to approximate the theoretical $E_{opt}$ according to the relation,
\begin{linenomath*}
\begin{equation}
f(E_i,\Delta E,q,t_c) = 0.87f_0(\Delta E,q). \label{eqn_Eidef}
\end{equation}
\end{linenomath*}
Here, the factor 0.87 follows from the simulation of the filling functions using the parameters in Table~\ref{tab:parametersimulation}. It can likewise be regarded as an empirical constant for the case at hand. For a Fermi-Dirac distribution (Eq. \ref{eqn_fepop_f0}), $E_i$ is related to $E_o$ (Eq. \ref{eqn_calEo}) by $E_i\approx E_o+2k_\textsc{b}T_\text{ch}$.
Therefore, the points $[E_i, f_0(\Delta E,q)]$ (blue dots in Fig. \ref{fig_fillFuncdE}c) that are obtained from the filling functions approximate the corresponding simulated but in practice unknown points $[E_{opt}, f_0(\Delta E,q)]$  (orange dots in Fig. \ref{fig_fillFuncdE}c). The magnitude function $f_m(E_t)$ can be obtained by interpolation and extrapolation of the experimental data sets $[E_i, f_0(\Delta E,q)]$ with desired $E_t$ resolution and range. According to the simulation, $E_i$ approaches $E_{opt}$ with relative uncertainty $<$ 5\% before $f_0(\Delta E,q)$ reaches $0.9\times \text{max} \{f_m(E_t)\}$. When $f_0(\Delta E,q )> 0.9\times \text{max} \{ f_m(E_t)\}$, replacing "0.87" in Eq. \ref{eqn_Eidef} with 0.98 will yield better results. \par

The second implication is that $f_m(E_t)$ is the envelope of the electron population functions derived from a uniform trap depth distribution ($N(E_t)=1$). From Fig. \ref{fig_fillFuncdE}c, for a given charging temperature $T_\text{ch}$, the filling function $f(E_t,\Delta E,q,t_c)$ is tangent to the magnitude function $f_m(E_t)$ in a small trap depth range which is illustrated in gray in the figure. This means that the envelope constructed from all electron population functions, as done in Sec. \ref{sec:n}, is the product of the envelope of filling function, which is $f_m(E_t)$, and an existing trap depth distribution $N(E_t)$. This underlies the method of extracting $N(E_t)$ that is elaborated in the coming paragraphs.

\subsubsection{Methods of extracting $N(E_t)$} \label{sec:2methods}
There are two methods to reconstruct the trap depth distribution $N(E_t)$. The first method relies on the magnitude function $f_m(E_t)$ and the envelope function $n_\text{env}(E_t,q,t_c)$. The trap depth distribution can be obtained according to
\begin{linenomath*}
\begin{equation}
N(E_t)=\frac{n_\text{env}(E_t,q,t_c)}{f_m(E_t)}. \label{eqn_extractN1}
\end{equation}
\end{linenomath*}
The second method originates from the idea of extracting $N(E_t)$ for the cases with $\Delta E=0.0 \text{ eV}$ aforementioned. At first, every electron population function $n(E_t,\Delta E,q,t_c)$ is  normalized by its magnitude function $f_0(\Delta E,q)$. In this way, every electron population function has the same magnitude of filling, meaning that the influence of $\Delta E$ has been removed. In the trap depth range $[E_o(\Delta E,q_1),E_o(\Delta E,q)]$, the value of trap depth distribution can be calculated by
\begin{linenomath*}
\begin{equation}
\frac{\Delta \tilde{n}_t(q,t_c)}{\delta E} = \frac{1}{\delta E}\int_{E_a}^{E_b}\Bigl[\frac{n(E_t,q_1,t_c)}{f_0(\Delta E,q_1)}-\frac{n(E_t,q,t_c)}{f_0(\Delta E,q)}\Bigr]dE_t 
\label{eqn_extractN2}
\end{equation}
\end{linenomath*}
in which $\delta E = E_o(\Delta E,q)-E_o(\Delta E,q_1),\text{ with }q\rightarrow T_\text{ch}$ and $q_1\rightarrow T_\text{ch}-\Delta T_\text{ch}$. The trap depth distribution $N(E_t)$ can be approximated by calculating $\Delta\tilde{n}_t(q,t_c)/\delta E$ for all available charging temperatures $T_\text{ch}$.

\subsection{Extraction from experiment}\label{sec:fexp}
As discussed in Sec. \ref{sec:analysisfilling}, the first step towards reconstructing $N(E_t)$ is to calculate the filling function $f(E_t,\Delta E,q,t_c)$. This function can be approximated by the following function in a relative manner,
\begin{linenomath*}
\begin{equation}
R(E_t,q,t_c)=\frac{n(E_t,q,t_c)}{n(E_t,q_r,t_c)}, \label{eqn_Rdef}
\end{equation}
\end{linenomath*}
in which the reference charging condition is $q_r\rightarrow T_\text{ch0}$. Here, $R(E_t,q,t_c)$ is thus termed the relative filling function.  For each $R(E_t,q,t_c)$, the magnitude $R_0(\Delta E,q)$ and the corresponding characteristic trap depths $E_o$ and $E_i$ can be extracted. The magnitude $R_0(\Delta E,q)$ is taken as the averaged $R(E_t,q,t_c)$ in a range where it has reached a plateau.  Oscillations in $R(E_t,q,t_c)$ bring uncertainties in the extraction. Therefore, an intermediate electron population function $n(E_t,q_r^\prime,t_c)$ can be used to calculate $R(E_t,q,t_c)$, meaning that
\begin{linenomath*}
\begin{equation}
R(E_t,q,t_c)=\frac{n(E_t,q,t_c)}{n(E_t,q_r^\prime,t_c)}\times R_0(\Delta E,q_r^\prime). \label{eqn_Rfuncref}
\end{equation}
\end{linenomath*}
Herein, the charging temperature $T_\text{ch}^\prime (\text{short for } q_r^\prime\rightarrow T_\text{ch}^\prime)$ can be chosen to be $\delta T$ smaller than $T_\text{ch}$. The value of $\delta T$, on the one hand should be small enough to reduce oscillations as much as possible since $n(E_t,q,t_c)$ suffers similar uncertainties to that of $n(E_t,q_r^\prime,t_c)$. On the other hand, $\delta T$ should be large enough to avoid distorting significantly the shape of the $R(E_t,q,t_c)$ in the $E_t$ range where $R(E_t,q,t_c)$ has not reached the plateau of $R_0(\Delta E,q)$. \par

\begin{figure}
\centering
\includegraphics[width = 0.98\linewidth]{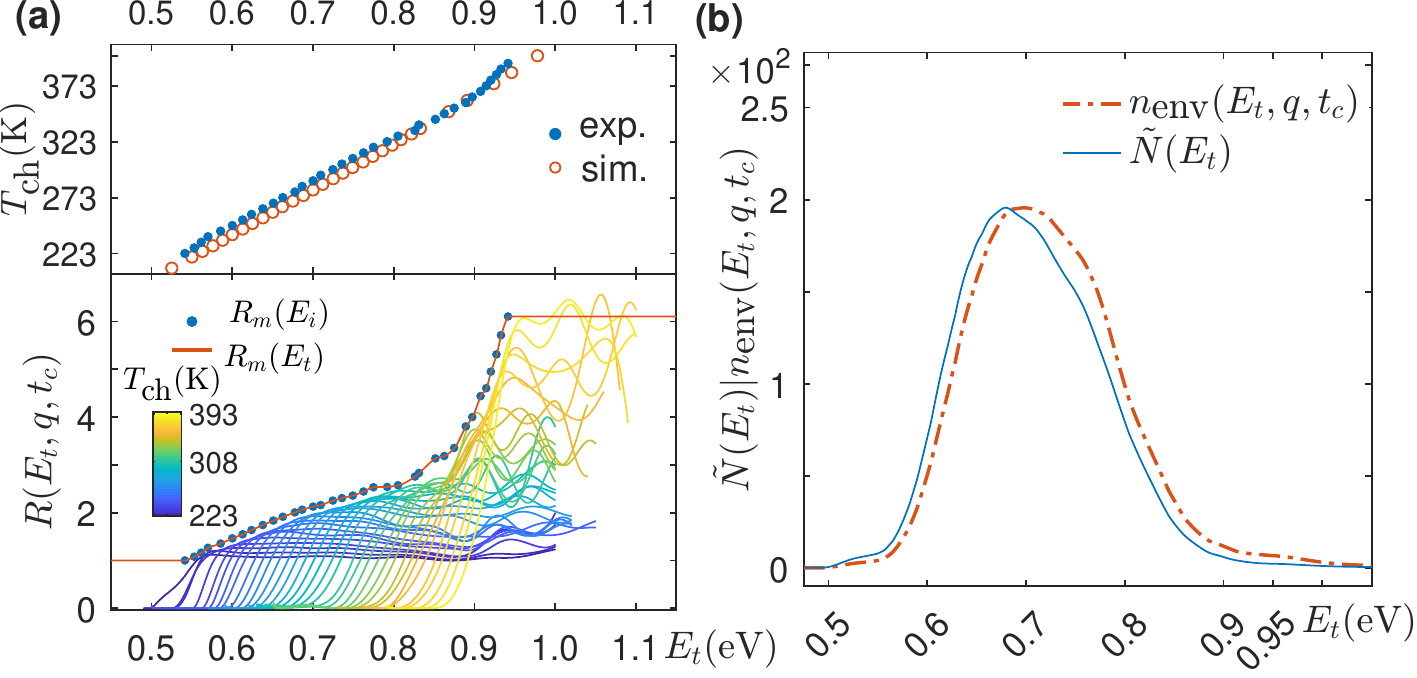} 
\caption{\textbf{Approximating the magnitude function}. \textbf{(a)} The relative filling function $R(E_t,q,t_c)$ (bottom panel, $q_r\rightarrow 223 \text{ K}$) approximates the filling function $f(E_t,\Delta E,q,t_c)$. The discrete $[E_i,R_0(\Delta E,q)]$ pair and the magnitude function $R_m(E_t)$ are displayed as blue dots and an orange line, respectively (bottom panel). The simulated $E_i$ agrees with the experimental ones within a constant difference (top panel). \textbf{(b)} The trap depth distribution $\tilde{N}(E_t)$ is calculated from the envelope $n_\text{env}(E_t,q,t_c)$ by using Eq. \ref{eqn_extractN1}. Note $n_\text{env}(E_t,q,t_c)$ has been scaled to the same magnitude of $\tilde{N}(E_t)$.} 
\label{fig_epopRatio} 
\end{figure}

After determining the magnitude $R_0(\Delta E,q)$, the characteristic trap depth $E_i$ and $E_o$ can be extracted according to Eq. \ref{eqn_calEo} and Eq. \ref{eqn_Eidef}, respectively.
The relative functions are shown in Fig. \ref{fig_epopRatio}a (bottom panel). After calculating $E_i$ and $R_0(\Delta E,q)$ for each relative filling function, an approximation to the magnitude function can be constructed: 
\begin{linenomath*}
\begin{equation*}
R_m(E_i): E_i(q)\mapsto R_0(\Delta E,q) \label{eqn_RmEi}
\end{equation*} 
\end{linenomath*}
which is a discrete analog to Eq. \ref{eqn_fmEtdef} (blue dots in Fig. \ref{fig_epopRatio}a, bottom panel). Here, $R_m(E_i)$ can be interpolated for $E_t\in[\text{min}(E_i),\text{max}(E_i)]$, and extrapolated beyond these limits by using $R_m\mathbf{(}\text{min}(E_i)\mathbf{)}$ and $R_m\mathbf{(}\text{max}(E_i)\mathbf{)}$ (Fig. \ref{fig_epopRatio}a). This leads to the approximated magnitude function $R_m(E_t)$ (orange line in Fig. \ref{fig_epopRatio}a, bottom panel), i.e.,
\begin{equation}
R_m(E_t): E_t(q)\mapsto R_0(\Delta E,q), \label{eqn_RmEt}
\end{equation}
in which $E_t$ is now in the full trap depth range of consideration, i.e., $E_t\in[E_a,E_b]$.
As shown in the top panel of Fig. \ref{fig_epopRatio}a, the experimental $E_i$ is almost linear with the charging temperature $T_\text{ch}$. The simulated $E_i$ differs from the experimental one by almost a constant amount for all $q\rightarrow T_\text{ch}$. The reason for the discrepancy will be discussed in Sec. \textsc{discussion}. \par

\subsubsection{Extracting $N(E_t)$ via Eq. \ref{eqn_extractN1}}
Since relative filling functions are used, the approximated trap depth $\tilde{N}(E_t)$ can be calculated by
\begin{equation}
\tilde{N}(E_t) = \frac{n_\text{env}(E_t,q,t_c)}{R_m(E_t)}, \label{eqn_extildeN1}
\end{equation} 
and the result is shown as blue line in Fig. \ref{fig_epopRatio}b. The shape of $\tilde{N}(E_t)$ differs slightly from the envelope $n_\text{env}(E_t,q,t_c)$. The resolution of the trap depth of this method is very high, and $R_m(E_t)$ only introduces relatively large uncertainties for $E_t>\text{max}(E_i)$ and $E_t<\text{min}(E_i)$ due to the extrapolation. However, the absolute uncertainties may be smaller because $N(E_t)$ has negligible value in these region. This can be further avoided by extending the range of charging temperatures. \par

\begin{figure}
\centering
\includegraphics[width = 0.98\linewidth]{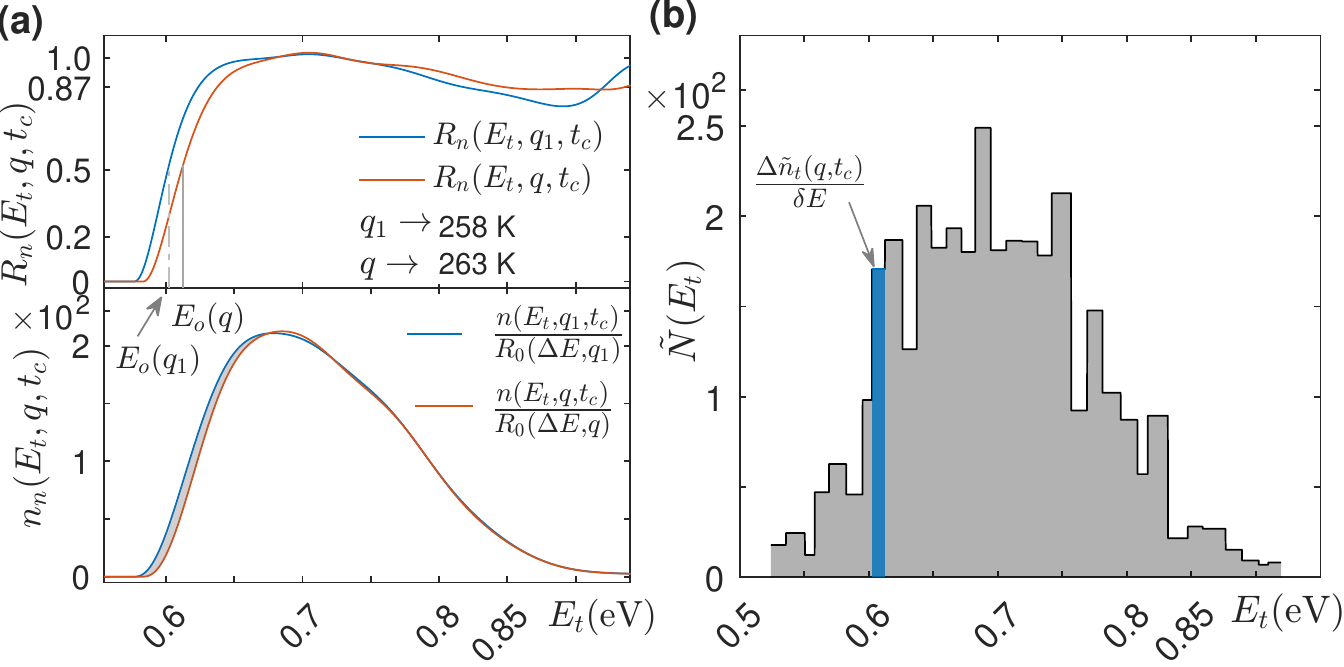} 
\caption{\textbf{Extracting $N(E_t)$ via Eq. \ref{eqn_extractN2}}. \textbf{(a)} The area between two normalized electron population functions for $q$ and $q_1$ characterizes $\Delta \tilde{n}_t(q,t_c)$ (gray area, bottom panel). Here, $R_n(E_t,q,t_c)=R(E_t,q,t_c)/R_0(\Delta E, q)$. \textbf{(b)} The trap depth distribution $\tilde{N}(E_t)$ according to Eq. \ref{eqn_extractN2} is represented by a histogram, together with  $\Delta \tilde{n}_t(q,t_c)/\delta E$ for $q\rightarrow 263 \text{ K}$ as a blue bar.} 
\label{fig_Koenmodified} 
\end{figure}

\subsubsection{Extracting $N(E_t)$ via Eq. \ref{eqn_extractN2}}
This method is easy to implement as it only requires discrete pairs of $E_o(q)$ and $R_0(\Delta E,q)$. For a pair of charging conditions $q\rightarrow T_\text{ch}$ and $q_1\rightarrow T_\text{ch}-\Delta T_\text{ch}$, the total number of trapped electrons in the range $[E_o(\Delta E,q_1),E_o(\Delta E,q)]$ can be calculated by Eq. \ref{eqn_extractN2} upon replacing $f_0(\Delta E,q)$ by $R_0(\Delta E,q)$. The gray area under the normalized electron population functions $n(E_t,q_1,t_c)/R_0(\Delta E,q_1)$ and $n(E_t,q,t_c)/R_0(\Delta E,q)$ in Fig. \ref{fig_Koenmodified}a (bottom panel) actually represents $\Delta \tilde{n}_t(q,t_c)$. The approximated trap depth distribution $\tilde{N}(E_t)$ is shown as a histogram in Fig. \ref{fig_Koenmodified}b, in which $\Delta \tilde{n}_t(q,t_c)$ has been added for illustration purposes. According to Eq. \ref{eqn_extractN2}, a large uncertainty in $n(E_t,q_1,t_c)/R_0(\Delta E,q_1)$ will lead to large uncertainties for two $\Delta \tilde{n}_t(q,t_c)$. This explains the occurrence of several pairs of high +low bin heights in the histogram.

\begin{figure}
\centering
\includegraphics[width = 0.98\linewidth]{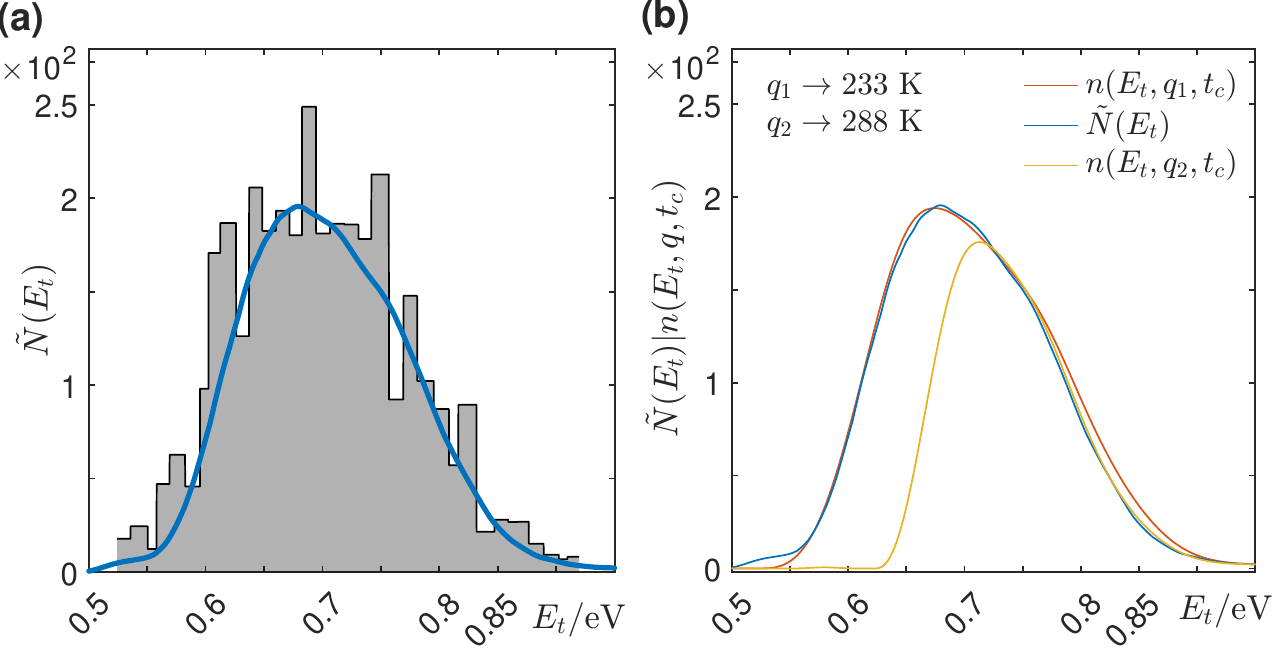} 
\caption{\textbf{Trap depth distribution of BaSi$_2$O$_2$N$_2$:2\%Eu$^{2+}$}. \textbf{(a)} The trap distribution $\tilde{N}(E_t)$ extracted by Eq. \ref{eqn_extildeN1} (blue) agree with that extracted from Eq. \ref{eqn_extractN2} (histogram). \textbf{(b)} The electron population function at low charging temperature, e.g., $q_1\rightarrow 233\text{ K}$, can approximate the trap depth distribution $N(E_t)$ but may fail in some ranges. } 
\label{fig_NEcompare} 
\end{figure}

The two methods above reach a consistent trap depth distribution $\tilde{N}(E_t)$ (Fig. \ref{fig_NEcompare}a). This validates the methods based on simulations in Sec. \ref{sec:analysisfilling}. The method of Eq. \ref{eqn_extildeN1} yields improved precision and resolution of $E_t$. It is noteworthy that an electron population function at low charging temperature, e.g., $q\rightarrow 233 \text{ K}$, can approximate the shape of $\tilde{N}(E_t)$ to a satisfactory extent (see $n(E_t,q_1,t_c)$ in Fig. \ref{fig_NEcompare}b, orange line). For higher charging temperature, only a part of the underlying trap depth can be revealed by the electron population function , e.g., $n(E_t,q_2,t_c)$ in Fig. \ref{fig_NEcompare}b (yellow line). The discrepancy between $\tilde{N}(E_t)$ and $n(E_t,q,t_c) (q\rightarrow 223 \text{ K})$  reveals possible error sources from the electron population function or the procedure of extracting trap depth distributions. Therefore, the trap depth distributions can be evaluated to the first-order approximation by the electron population function $n(E_t,q,t_c)$ with the lowest possible charging temperature if the signal strength of the TL curve is still strong enough for the Tikhonov regularization process. This is beneficial for fast screening of persistent phosphors based on their trap depth distributions.

\subsubsection{The influence of the frequency factor $\nu_r$}
\begin{figure}[!b]
\centering
\includegraphics[width = 0.98\linewidth]{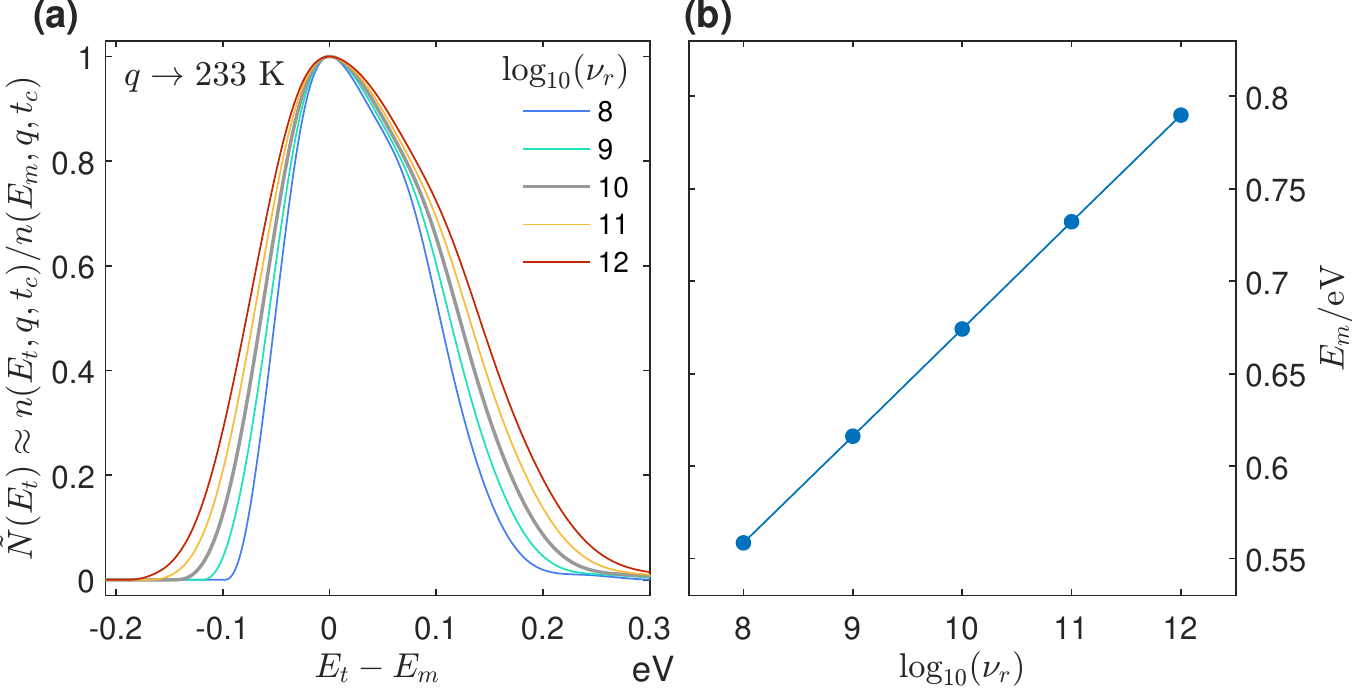} 
\caption{\textbf{The influence of the recombination frequency factor $\nu_r$}. \textbf{(a)} Changing the frequency $\nu_r$ to $a\nu_r$ by multiplying a positive scalar $a$ will compress ($a<1$) or stretch ($a>1$) the trap depth distribution and shift it along the $E_t$ axis. \textbf{(b)} The value of trap depth $E_m$ increases almost linearly with increasing $\textrm{log}_{10}[\nu_r(\text{s}^{-1})]$. Note the trap depth distribution can be approximated by the normalized electron population function $n(E_t,q,t_c)/n(E_m,q,t_c)(q\rightarrow 233\textrm{ K})$, with $n(E_m,q,t_c) \text{ and } E_m$ being the maximum of $n(E_t,q,t_c)$ and the corresponding trap depth, respectively.} 
\label{fig_Influencenur} 
\end{figure}

The current model assumes a fixed frequency factor of $\nu_r=10^{10}$ s$^{-1}$. Scaling $\nu_r$ by a positive factor $a$ will compress ($a<1$) or stretch ($a>1$) the trap depth distribution and shift it along the $E_t$ axis (Fig. \ref{fig_Influencenur}). As shown in previous sections, the trap depth distribution can be approximated by a normalized electron population function, i.e., $\tilde{N}(E_t)\approx n(E_t,q,t_c)/n(E_m,q,t_c)(q\rightarrow 233\textrm{ K})$. Here, $n(E_m,q,t_c) \text{ and } E_m$ are the maximum of $n(E_t,q,t_c)$ and the corresponding trap depth, respectively. The value of $E_m$ characterizes the position of the trap depth distribution along the $E_t$ axis. It is obvious that $E_m$ scales almost linearly with $\textrm{log}(\nu_r)$ (Fig. \ref{fig_Influencenur}b). Hence, small deviation from the chosen frequency factor hardly imposes significant impact on the trap depth distribution. 

\section{Discussion}
\subsection{Interpretation of the model}
\subsubsection{Approximating the kernel}
The temperature integral Eq. \ref{eqn_FBalarin} decreases almost exponentially with decreasing temperature $T$. The influence of $F(E_t,T_0)$ on the kernel $K(E_t,T)$ (Eq. \ref{eqn_kernelEtT}) is mainly located at low temperature (high values in the color plot in Fig. \ref{fig_TLepop}). The term $F(E_t,T_0)$ is thus sufficiently smaller than $F(E_t,T)$ and can be neglected in Eq. \ref{eqn_kernelEtT} when $T$ is about 20-30 K greater than $T_0$ for traps that are sufficiently deep. The kernel $K(E_t,T)$ now reads,
\begin{linenomath*}
\begin{subequations} \label{eqn_Kernelsimple}
\begin{align}
\begin{split}
K(E_t,T) =& \frac{W(\text{e}\nu_rT/\beta)}{T} \text{exp}\Bigl[-\frac{E_t-E_s(T)}{k_\textsc{b}T} \\
& -\text{exp}\Bigl(-\frac{E_t-E_s(T)}{k_\textsc{b}T}\Bigr)\frac{\bigl(E_s(T)+k_\textsc{b}T\bigr)/E_t}{\sqrt{1+4k_\textsc{b}T/E_t}}\Bigr], 
\end{split} \label{eqn_KernelEs}
\\
E_s(T) =& k_\textsc{b}T\left[W(\text{e}\nu_rT/\beta)-1\right], \label{eqn_Esdef}
\end{align}
\end{subequations}
\end{linenomath*}
where $W(x)$ is the Lambert function of the 0$^\text{th}$ branch. (The derivation is given in Sec. VI in the SM \cite{SM}.)  Here, $E_s(T)$ refers to the trap depth that corresponds to the maximum of the kernel. When $E_s(T)$ is several $k_\textsc{b}T$ smaller than $E_t$, the kernel Eq. \ref{eqn_Kernelsimple} can be further simplified as,
\begin{equation*}
 K(E_t,T)\approx \frac{\nu_r}{\beta}\text{exp}\left(-\frac{E_t}{k_\textsc{b}T}\right),
\end{equation*} 
which underlies the initial rise method. The implicit assumption means that the extracted trap depth will be underestimated up to several $k_\textsc{b}T$. The magnitude of the kernel $K(E_t,T)$, which is ${W(\text{e}\nu_rT/\beta)}/{T}$, decays with increasing temperature $T$. The shape of $K(E_t,T)$ is close to the probability density function (\textsc{pdf}) of the Gumbel distribution \cite{coles_introduction_2001}, and the standard deviation is proportional to $k_\textsc{b}T$. This means the kernel is mainly distributed several $k_\textsc{b}T$ around $E_s$ and a linear band structure can be found in the discrete $K(E_t,T)$ (Fig. \ref{fig_TLepop}a). Hence, the information of the electron population function gets more smeared out in TL curves when temperature increases, which is one of the reasons to apply the Tikhonov regularization method. 

At large argument $x$, the derivative $W'(x)=[x+\text{exp}(W(x))]^{-1}$ is close to zero. Hence, the Lambert function in Eq. \ref{eqn_Esdef} can be replaced by its averaged value $\langle W\rangle$ in a reasonable temperature range (e.g., 100 to 600 K), and Eq. \ref{eqn_Esdef} becomes
\begin{linenomath*}
\begin{equation}
E_s(T) \approx (\langle W\rangle-1) k_\textsc{b}T
\end{equation}
\end{linenomath*}
This clearly indicates a linear relationship between temperature $T$ and the characteristic trap depth $E_s(T)$, which has been shown in Fig. \ref{fig_TLepop}a. For a delta distribution $N_0\delta(E_t-E_0)$, the trap depth $E_0$ can be estimated from the maximum of the TL glow curve, i.e., $T_m$, via Eq. \ref{eqn_Esdef}. For a fixed $\nu_r/\beta = 10^{9}$, the Urbach relation is recovered, i.e. $E_s(T_m) \approx 23.3 k_\textsc{b}T_m\approx T_m/500$ \cite{Ubach1930Zur}.\par

A special case of TL is isothermal thermoluminescence, known as persistent luminescence (PersL) or afterglow. After charging the phosphor at $T_\text{ch}$, the temperature remains unchanged while the luminescence intensity $I(t_0)$ is recorded as a function of delay time $t_0$, i.e., the decay profile of the PersL is obtained. The PersL intensity $I(t_0)$ can also be written as the integral equation,
\begin{linenomath*}
\begin{equation}
I(t_0)=\int_0^{\infty}n(E_t,q,t=0)K(E_t,t_0)dE_t,\label{eqn_PersLIt0}
\end{equation}
\end{linenomath*}
with the kernel,
\begin{linenomath*}
\begin{subequations}
\begin{align}
\begin{split}
K(E_t,t_0)=&\frac{1}{t_0}\text{exp}\Bigl[-\frac{E_t-E_s(t_0)}{k_{\textsc{b}}T_\text{ch}}\\
&\qquad\quad -\text{exp}\Bigl(-\frac{E_t-E_s(t_0)}{k_{\textsc{b}}T_\text{ch}}\Bigr)\Bigr]
\end{split} \label{eqn_GumbelPersL} \\
E_s(t_0)&=k_{\textsc{b}}T_\text{ch}\text{ln}(\nu_rt_0), \label{eqn_Es4AG}
\end{align}
\label{eqn_PersLkernel}
\end{subequations}
\end{linenomath*}
which follows the \textsc{pdf} of the Gumbel distribution \cite{coles_introduction_2001}.
This immediately indicates the presence of the power law of $t_0^{-\alpha}$ ($\alpha\approx$ 1) for the decay profile, which has been observed in many persistent phosphors \cite{MedlinPhysRev,Huntley_tunneling}. In theory, trap parameters can also be inferred from the PersL decay profile $I(t_0)$ \cite{chen_TLPersLDelay_1986,ChenMcKeever97book}. The shape of the kernel $K(E_t,t_0)$ remains unchanged, but its magnitude ${t_0}^{-1}$ will scale down the light output from deep traps significantly. This requires both huge delay time $t_0$ to probe a wide range of trap depths and highly sensitive detectors with a large dynamic range to register $I(t_0)$ out of noises. However, the decay profile $I(t_0)$ can be used in conjunction with TL curves to understand materials properties to a deeper extent \cite{InvsProbJPCA,GarnetAG_trapdistrib}. \par

\subsubsection{Frequency factor $\nu_r$}
The frequency factor $\nu_r$ has been fixed to $10^{10}$ s$^{-1}$ for regularization in this paper. There are methods to extract the frequency factor, for example see Refs. \cite{RandallWilkins_I,detfreqfactor_OptMater},  but they are obtained under the assumption of one discrete trap depth. Recently, McKeever and Sholom proposed a more sophisticated method of extracting both distributed trap depths and frequency factors \cite{EtnuMccKeever21}. The precise determination asks for detailed knowledge of TL and PersL, and carefully designed experiments. 

Here, we consider the effect of scaling $\nu_r$ by a positive scalar $a$ on the obtained trap depth distribution, which has been illustrated in Sec. \ref{sec:fexp}. The Lambert function is expanded as,
\begin{linenomath*}
\begin{equation*}
W(x) \approx \text{ln}(x)-\text{ln}\left[\text{ln}(x)\right], \label{eqn_Wapprox}
\end{equation*}  
\end{linenomath*}
at large values for $x$ \cite{LambertW_approx}.
Hence, the characteristic trap depth can be approximated as,
\begin{linenomath*}
\begin{equation*}
E_s(T,a\nu_r) \approx E_s(T,\nu_r)+\text{ln}(a)k_\textsc{b}T.
\end{equation*}
\end{linenomath*}
This shows that the extracted electron population function can be scaled along $E_t$ due to the term $\textrm{ln}(a)$. Furthermore, the position of the electron population function, which was estimated by $E_m$ (Sec. \ref{sec:fexp}), will be shifted by an amount that is proportional to $\textrm{ln}(a)$. The extracted trap depth distribution can be altered similarly since it can be approximated by an electron population function for sufficiently low charging temperature (e.g., $n(E_t,q,t_c)$ with $q\rightarrow 233 \textrm{ K}$). \par 

\subsubsection{The optimal trap depth at charging temperature $T_{ch}$}
The linear relationship between the charging temperature and the trap depth which can be optimally charged at that temperature (see Fig. \ref{fig_fillFuncdE}) can be understood to a satisfactory extent. We set $\sigma_\text{osl}$ to zero without loss of generality. At a given charging condition $q$, the magnitude of filling,
\begin{linenomath*}
\begin{equation}
\begin{split}
&f_0(\Delta E,q)  =  \frac{k_\text{trap}(\Delta E,q)}{k_\text{trap}(\Delta E,q)+k_\text{rcb}(E_t,q)} \\
 & = {\left[1+\frac{A_t\nu_rk_\text{rad}}{(A_t+\nu_r)\nu_t\sigma_\text{abs}I_e(\lambda)}\text{exp}\left(-\frac{E_t-\Delta E}{k_\textsc{b}T}\right)\right]}^{-1},
\end{split}
\end{equation}
\end{linenomath*}
increases with increasing $E_t$. Meanwhile, the remainder of Eq. \ref{eqn_fillt0}:
\begin{linenomath*}
\begin{equation*}
1-\text{exp}\left[-\left(k_\text{trap}(\Delta E,q)+k_\text{rcb}(E_t,q)\right)t_\text{ch}\right],
\end{equation*}
\end{linenomath*}
decreases with increasing $E_t$. This leads to a trap depth at which the phosphor can be charged to the largest efficiency at the given charging temperature $T_\text{ch}$ (Fig. \ref{fig_fillFuncdE}a, bottom panel). This sets the relationship between $T_\text{ch}$ and $E_{opt}$. It is interesting to note that the magnitude $f_0(\Delta E,q)$ shows an effective activation energy of $E_t-\Delta E$ to 100\% filling. \par

\subsubsection{First-order kinetics}
We now turn to the first-order kinetics, which determines the validity of the methods. The first-order kinetics, which originates from the isolated-pair approximation without considering retrapping (Sec. \ref{sec: localmodel}), has been assumed for both detrapping during TL and the filling of traps during charging. Noticeably, retrapping has an impact when the density of electrons at the excited state of traps or of luminescent activators is increased significantly by optical stimulation \cite{ZnSiOTunneling,ZWPanSrMgGeOPbOSLdetrapping} or even mechanical stimulation \cite{AddingMemory}, enhancing the probability of trapping for these electrons.  The isolated-pair approximation actually implies that the density of electrons at the excited state of traps or of luminescent activators is small enough (see Sec. \ref{sec: localmodel}). Furthermore, the experimental conditions can be carefully designed to minimize the effect of retrapping. According to the charging procedure (see Sec. \ref{sec:matermethod}), the  phosphor was charged at high irradiance and long charging duration to reach saturated electron population functions. In this way, non-first-order kinetics during charging can be smeared out into the thermal equilibrium. The cooling process (to $T_0$) after charging further reduces the non-first-order kinetics. Hence, the limited rate of retrapping process is not likely to pose a large impact on the methods of extracting the trap depth distribution. 

First-order kinetics of electrons among traps with a trap depth distribution can induce the shape of the electron population evolves with increasing charging duration (with fixed charging irradiance). For the model phosphor BaSi$_2$O$_2$N$_2$:2\%Eu$^{2+}$, the experimental observation and simulation reveal that the trap depth at the maximum of the electron population function $E_m$ increases before reaching a plateau as the charging duration increases (see Sec. V in SM \cite{SM}). It is often exploited that it follows from first-order kinetics that the shape of the TL curve and the resultant shape of the electron population function are independent of the charging duration (charging irradiance fixed). If deviations occur, this is usually interpreted as the result of non-first-order kinetics \cite{e_trapmechanism_ProcPhysSoc}. However, this only applies for phosphors with just one discrete trap depth. This suggests that first-order kinetics can be more dominant in persistent and storage phosphors than expected, and thus re-trapping in TL can be safely ignored accordingly.

\subsection{Analysis of uncertainties} \label{sec:uncertaintyAnalysis}
The accuracy of the extracted trap depth distribution relies on the theoretical framework that suggests the methods for extracting information. The most important implication of the first-order kinetics of charging is the presence and scientific significance of the magnitude function $f_m(E_t)$, Eq. \ref{eqn_fmEtdef} (or the relative version $R_m(E_t)$, Eq. \ref{eqn_RmEt}). On the one hand, $f_m(E_t)$ is the magnitude that should be used to correct electron population functions for variable charging temperature to remove the influence of the thermal barrier for charging $\Delta E$. This directly results in the method via Eq. \ref{eqn_extractN2}. On the other hand, $f_m(E_t)$ is also the envelope of the electron population functions at variable $T_\text{ch}$ originating from a uniform trap depth distribution $N(E_t)=1$. Evidently, the trap depth distribution can be recovered by using the envelope $n_\text{env}(E_t,q,t_c)$ and the magnitude function $f_m(E_t)$ according to Eq. \ref{eqn_extractN1}. These deductive methods do not depend on the parameters used in the simulation but depend on the presence of the thermal barrier $\Delta E$. The effect of $\Delta E$ can be canceled out via Eq. \ref{eqn_extractN1} or Eq. \ref{eqn_extractN2} without knowing its exact value. Systematic errors are thus minimized, and random errors originate from the calculation of the electron population functions and the associated magnitude function directly. \par

Consistent experimental settings for charging should be guaranteed as much as possible. It is advised to cool the phosphor after charging to $T_0$ at a fast cooling rate such that the electron population function $n(E_t,q,t_c)$ ($q\rightarrow T_0$) can be best approximated by  $n(E_t,q,t=0)$ ($q\rightarrow T_\text{ch}$). The charging irradiance $I_e(\lambda)$ and charging duration $t_\text{ch}$ should be sufficiently large to produce a high signal strength and also a stable shape of the electron population functions. (A detailed analysis of dose dependency can be found in Sec. V of SM \cite{SM}.) The noise level should be minimized in order to increase the signal-to-noise ratio which is important to generate high-quality electron population functions via the Tikhonov regularization. 

Numerical uncertainties mainly originate from the relative filling function $R(E_t,q,t_c)$ and the methods to extract the magnitude $R_0(\Delta E,q)$ and the characteristic trap depths $E_o$ (Eq. \ref{eqn_NOdEHeavyside}) and $E_i$ (Eq. \ref{eqn_Eidef}). The regularization method can yield small oscillations in the electron population functions because it uses oscillatory singular vectors to reconstruct solutions. The oscillation in $R(E_t,q,t_c)$ can be reduced by choosing an optimized reference electron population function (Eq. \ref{eqn_Rfuncref}), leading to a more reliable magnitude $R_0(\Delta E,q)$. The characteristic trap depth $E_i$ from Eq. \ref{eqn_Eidef} will yield a few percent of deviation from the model. Large uncertainties may arise when there is an oscillation of $R(E_t,q,t_c)$ before it reaches the magnitude. This has been shown for $E_i\in (0.8,0.9) \text{ eV}$ in Fig. \ref{fig_epopRatio}.

The method of extracting trap depth distributions via Eq. \ref{eqn_extractN1} suffers from uncertainties originating from both $R_m(E_t)$ and the envelope of the electron population functions, $n_\text{env}(E_t,q,t_c)$. Meanwhile, the method via Eq. \ref{eqn_extractN2} is prone to errors in $n_t(q,t_c)$ and the magnitude $R_0(\Delta E,q)$. As the method of difference is used, a large uncertainty in $n_t(q,t_c)/R_0(\Delta E,q)$ will definitely produce large uncertainties in $N(E_t)$ in two consecutive trap depth ranges. This can be confirmed by several pairs of "high+low" height of bins in the histogram of $\tilde{N}(E_t)$.

\subsection{Application of the method}
The present model assumes the presence of only one luminescent activator and only one thermal barrier. Real persistent phosphors can have multiple luminescent activators and traps or one kind of luminescent activator at multiple crystallographic sites, each providing its own emission spectrum and thermal barrier. In these cases, the first-order kinetics must be applied to each distinctive trapping-recombination process independently and the output is the sum of these independent processes. Ideally, spectrally resolved  recording of the TL intensity allows to distinguish the contributions of independent recombination processes. \par

Trap depth distributions have already found their way to technological applications. The obvious one is to understand and tune the PersL behavior of persistent phosphors. For example, the trap depth distribution in garnet phosphors can be tuned by alloying to optimize optical storage properties \cite{YAGtrap}. Furthermore, it provides an estimate for the optimum charging and working temperature ($T_{opt}$) of persistent phosphors.  The quantity $I(t_0)t_0$ can be used to quantify the luminescence decay profile. It combines the effect of intensity of afterglow and the noise level. According to Eq. \ref{eqn_PersLkernel}, the kernel $K(E_t,t_0)$ can be approximated by a boxcar function $\text{rect}\left(\frac{E_t-E_s(t_0)}{\pi k_\textsc{b}T_\text{ch}/\sqrt{6}}\right)$, leading to, 
\begin{linenomath*} 
\begin{equation}
I(t_0)t_0 \approx \frac{\pi}{\sqrt{6}}k_\textsc{b}T_\text{ch}\times n(E_s(t_0),q,t=0)
\end{equation}
\end{linenomath*}
which clearly indicates the influence of charging temperature and the trap depth distribution. The charging temperature $T_\text{ch}$ that maximizes $I(t_0)t_0$ can be estimated by examining the maximum of $N(E_t)$.  For BaSi$_2$O$_2$N$_2$:2\%Eu$^{2+}$, this optimum charging temperature is $\sim 288$ K, which results in an electron population function with its maximum located around that of the trap depth distribution (yellow line, Fig. \ref{fig_NEcompare}b). This prediction can be compared with further experimental verification.  

Given the trap depth distribution $\tilde{N}(E_t)$, PersL decay profiles or TL glow curves can be straightforwardly simulated at any charging and working conditions. This helps to explain and predict the properties of phosphors.  More importantly, the trap distribution can be used as a reliable feature of persistent and storage phosphors. This facilitates the discovery of empirical laws that govern the properties of persistent phosphors, e.g., via machine learning. \par

\section{Conclusion}
In this paper, a method was proposed to extract the trap depth distribution from thermoluminescene (TL) curves with the presence of a thermal barrier for charging. It is based on a local model for trapping and recombination that leads to first-order kinetics. The model predicts the evolution of the filling function as a function of charging temperature. In the first step of the method, the electron population functions $n(E_t,T_\text{ch},t_c)$ and the envelope $n_\text{env}(E_t,q,t_c)$ were obtained from the corresponding TL curves by the Tikhonov regularization method. In the second step, the relative magnitude of the filling function, i.e., $f_m(E_t)$, is estimated out of ratios of electron population functions.  Finally, the trap depth distribution can be estimated according to either ${N}(E_t)=n_\text{env}(E_t,q,t_c)/f_m(E_t)$ (Eq. \ref{eqn_extractN1}) or Eq. \ref{eqn_extractN2}. The methods do not require the value of the thermal barrier $\Delta E$ beforehand, although $\Delta E$ influences the filling functions. Our case study on \ce{BaSi2O2N2}:\ce{Eu^2+} validated this method. A broad trap depth distribution, ranging from 0.5 to 0.9 eV with the maximum $\sim 0.65$ eV, was revealed, assuming a frequency factor of $\nu_r=10^{10}$ s$^{-1}$. 

The method via Eq. \ref{eqn_extractN1} not only shows a clear physics picture but also yields high precision and resolution of trap depth, provided that TL curves with high signal strength and high signal-to-noise ratio are available. The trap distribution definitely promotes the understanding and tailoring of the properties of persistent and storage phosphors.
\begin{acknowledgments}
A.F. and P.F.S acknowledge the financial support of the Special Research Fund (BOF) via the GOA-"Enclose" project from Ghent University. J.J.J acknowledges the UGent Special Research Fund (Grant No. BOF/PDO/2017/002101). J.D. acknowledge the support by the BOF postdoctoral fellowship (No. BOF20/PDO/015) of Ghent University and the Natural Science Foundation of Jiangsu Province (BK20210481), China. 
\end{acknowledgments}



%

\bibliography{PhysRevB.bib}
\clearpage

\title{Supplemental Material:\\Revealing trap depth distributions in persistent phosphors with a thermal barrier for charging}

%

\date{\today}
\begin{abstract}
It covers supplementary information on experimental methods, extra experimental data, derivation of formulae and \textsc{matlab} codes that implement the regularization method.
\end{abstract}

\renewcommand{\thefigure}{S\arabic{figure}}
\renewcommand{\thetable}{S\arabic{table}}
\renewcommand{\theequation}{S\arabic{equation}}

\maketitle
\setcounter{section}{0}
\setcounter{figure}{0}
\setcounter{equation}{0}

\section{Thermal quenching profiles}
To obtain thermal quenching (TQ) profiles, the phosphor is heated from low temperature to high temperature continuously at a fixed rate while being illuminated by excitation light. The integrated intensity of the emission spectra as a function of temperature is the so-called TQ profile. Electron trapping in persistent or storage phosphors can reduce the emission intensity, therefore we adopt the method from Ref. \cite{Anorthite}.
\begin{figure}
\begin{center}
\includegraphics[width = 0.98\linewidth]{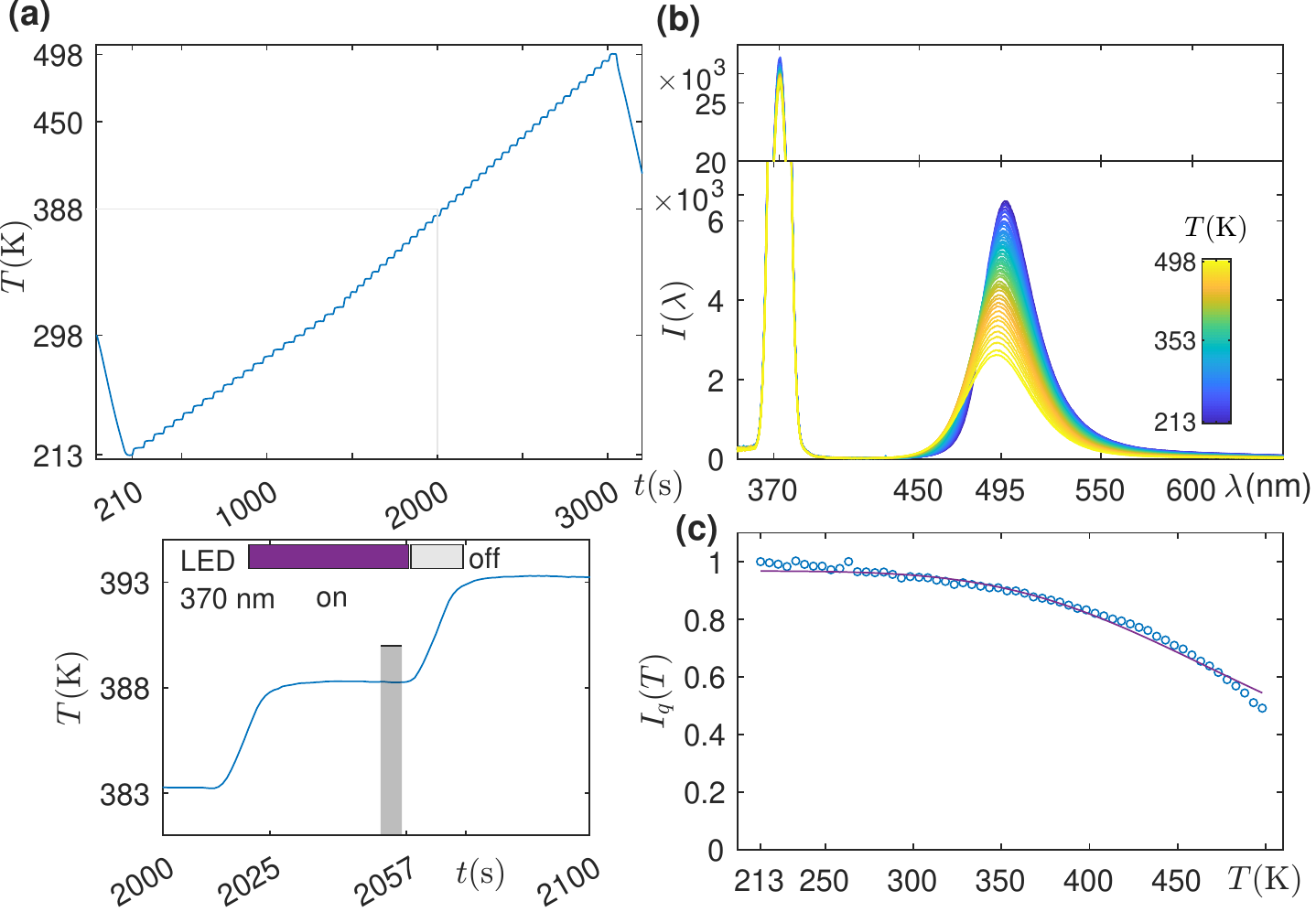} 
\caption{\textbf{Thermal quenching profiles.} \textbf{(a)} The phosphor was warmed up from 213 to 498 K with charging of 30 s at each temperature $T=213+5i$ K ($i=[0:1:57]$). Five spectra from the  24-28$^{\text{th}}$ s was averaged to $I(\lambda)$ (illustrated as a gray band in the bottom panel). \textbf{(b)} With increasing $T$, the intensity of the spectra of excitation light decreases only slightly while that of the emission spectra of \ce{BaSi2O2N2}:2\%\ce{Eu^2+} decreases significantly. \textbf{(c)} The integrated emission intensity (400-650 nm) was normalized to that of $T=$ 213 K, and was then fitted to the single barrier model $I_q(T) = {I_0}/\left[1+A \textrm{exp}\left(-\frac{E_q}{k_{\textsc{b}}T}\right)\right]$. The thermal barrier is found to be $E_q=0.254$ eV.}
\label{fig_TQall} 
\end{center}
\end{figure} 

The charging protocol (Fig. \ref{fig_TQall}a) characterizes charging  at each $T=213+5i$ K ($i=[0:1:57]$) for 30 s during heating from 213 to 498 K. The emission spectrum for each charging temperature $T$ was obtained by averaging five spectra from the 24-28$^{\text{th}}$ s of the charging, as shown by the gray band in Fig. \ref{fig_TQall}a. Each of these spectra (Fig. \ref{fig_TQall}b) was integrated from 400 nm to 650 to calculate the emission intensity $I_t(T)$. Then, $I_t(T)$ was normalized with respect to $I_t(213\textrm{ K})$, leading to the TQ profile $I_q(T)$ (Fig. \ref{fig_TQall}c).

The TQ profile (Fig. \ref{fig_TQall}c) was fitted to the single-barrier model \cite{TQ_Mott},
\begin{equation}
I_q(T)=\frac{I_0}{1+A\textrm{exp}\left(-\frac{E_q}{k_{\textsc{b}}T}\right)},
\end{equation}
where $k_{\textsc{b}}$ is the Boltzmann constant. $I_0, A$ and $E_q$ are fitting parameters. The results of the fit are $I_0=0.9676$, $A=289.4$ and $E_q=0.254$ eV. 

\section[The regularization method]{Extracting electron population function}
In the framework of first-order kinetics, the electron population function $n(E_t,q,t_c)$ is related to the TL intensity $I(t)$ via Eq. 4 in the paper. The first step to solve these equation is to discretize the integral equation over a grid $[T_0,T_m]\times[E_a,E_b]$. The quadrature method with the midpoint rule yields \cite{Hansen_reguBook},
\begin{equation}
\sum_{j=1}^N \omega_jK\left({E_t}_j,T_k\right)n\left({E_t}_j\right)=I\left(T_k\right), \quad k=1,2,...,M
\end{equation}
or in the matrix form (Eq. 6 in the paper),
\begin{equation}
Kn=I, \label{eqn_KniSM}
\end{equation}
with $K_{kj}=\omega_jK\left({E_t}_j,T_k\right), n_j=n\left({E_t}_j\right), I_k = I\left(T_k\right), \text{and } \omega_j=\frac{E_b-E_a}{N}$. Note that $I$ is obtained by interpolating the experimental TL data onto the temperature vector \texttt{t = T0+(0.5:1:M-0.5)'*dT}, with \texttt{dT=(Tm-T0)/M}. It is not possible to solve Eq. \ref{eqn_KniSM} via the standard least squares method due to the noise of TL signal and a huge condition number of $K(E_t,T)$ \cite{Hansen_reguBook}. The Tikhonov regularization method boils down to solve,
\begin{equation}
\begin{bmatrix}
 K\\ 
\lambda L
\end{bmatrix}n=\begin{bmatrix}
 I\\
0
\end{bmatrix}, \label{eqn_KLni}
\end{equation}
with an optimized regularization parameter $\lambda$ that minimizes the functional,
\begin{equation}
V(\hat{n})=\lVert K\hat{n} - I\rVert _2^2+ {\lambda}^2 \lVert L\hat{n}\rVert_2^2.
\end{equation} 
Herein, $L$ is the discrete approximation of a derivative operator. 
\begin{figure}
\begin{center}
\includegraphics[width = 0.80\linewidth]{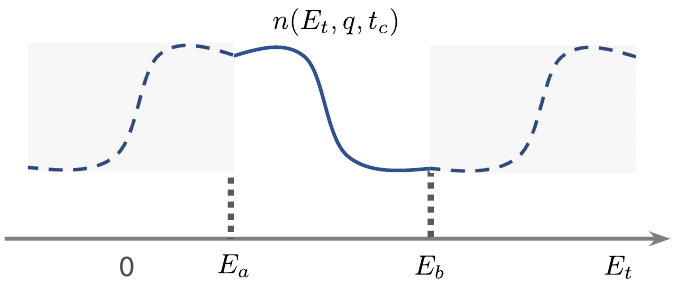} 
\caption{\textbf{Reflexive boundary condition.} The reflexive boundary condition assumes the solution $n(E_t,q,t_c)$ outside the interval $(E_a, E_b)$ results from reflecting $n(E_t,q,t_c)$ for $E_t\in (E_a,E_b)$ along $E_t = E_a$ and $E_t = E_b$.}
\label{fig_reflexBC} 
\end{center}
\end{figure} 

To solve this problem, reflexive boundary conditions are adopted, meaning that $n(E_t,q,t_c)$ for $E_t<E_a$ and $E_t>E_b$ result from reflecting $n(E_t,q,t_c)$ along $E_t = E_a$ and $E_t = E_b$, respectively (Fig. \ref{fig_reflexBC}). Under such a condition, the kernel is now,
\begin{equation}
\begin{split}
{K\left(E_t,T\right)}_r = & K\left(E_t,T\right)\\
&+K\left(2E_a -E_t,T\right)\\
&+K\left(2E_b -E_t,T\right),
\end{split} \label{eqn_Kernelr}
\end{equation}
subjected to $K\left(2E_a -E_t,T\right)=0$ when $2E_a -E_t<0$. Furthermore, the discrete approximation of a second derivative operator $L$ can be written as \cite{Hansen_reguBook,TikhReguBC},
\begin{equation}
L=\begin{bmatrix}
 -1 &  1&  &  & \\
  1&  -2&  1&  & \\
  &  .&  .&  .& \\
  &  &  1&  -2& 1\\
  &  &  &  1& -1
\end{bmatrix}_{N\times N}.
\end{equation}
A generalized singular value decomposition (GSVD) method is utilized to decompose $K$ and $L$ simultaneously so that the solution is given by,
\begin{equation}
n_{L,\lambda} = \sum_{k=1}^N\phi_{k}^{[L,\lambda]}\frac{{u_k^{\prime}}^\textsc{t}I}{\sigma_k^{\prime}}n_k^{\prime}, \label{eqn_xGSVD}
\end{equation}
where the ratios $\sigma_k^{\prime}/\mu_k^{\prime}$ (with ${\sigma_k^{\prime}}^2+{\mu_k^{\prime}}^2=1$) are the generalized singular values. The right singular vectors $n_k^{\prime}$, which are shared by both $L$ and $A$, are mutually independent but are neither normalized nor orthogonal. There are two sets of left GSVD vectors, $u_k^{\prime}$ and $v_k^{\prime}$, that are mutually orthonormal and that satisfy
\begin{equation*}
Kn_k^{\prime}=\sigma_k^{\prime}u_k^{\prime}, \quad Ln_k^{\prime}=\mu_k^{\prime}v_k^{\prime}.
\end{equation*}
Here we provide the \textsc{matlab} code that solves $n(E_t,q,t_c)$ from TL data by using the \textit{Regularization Tools} \textsc{matlab} package \cite{RegToolsRef0,RegToolsRef}. This package needs to be added to the search path of \textsc{matlab} installation if following code is to be used.  
\begin{itemize}
\item Discretization. 
\begin{small}
\begin{lstlisting}[style = Matlab-editor]
kB = 0.08617; nu = 1e10; beta = 0.5; 
T0 = 213.15; Tm = 473.15; %T range, supplied by the user.
Ea = 300; Eb = 1200; % Et range, meV, specified by the user.
N = 3200; M = ceil(N/2); % # of intervals
de = (Eb-Ea)/N; dT = (Tm-T0)/M;   
e = Ea+(0.5:N-0.5)'*de; t = T0+(0.5:M-0.5)'*dT; 
[E,T] = meshgrid(e,t); % meshgrid on the EtxT plane;
tInt = @(x,y) kB*nu/beta*(y.^2./x).*...
exp(-x./y/kB)./sqrt(1+4*kB*y./x); % x--Et, y--T; Ref. M Balarin J therm Anal 12, 169 (1977).
TempInt = @(x,y) tInt(x,y)-tInt(x,T0);% temperature integral
Kf = @(x,y) de*nu/beta*exp(-x./(kB*y)-TempInt(x,y));
K1 = Kf(E,T); K2 = Kf(2*Eb-E,T);
K3 = Kf(2*Ea-E,T); 
zIdx = (2*Ea-E)<0; K3(zIdx) = 0; 
K = K1+K2+K3;
L = diag([-1;ones(N-2,1)*(-2);-1]);
L(2:N,1:N-1) = L(2:N,1:N-1) +...
 diag(ones(N-1,1));
L(1:N-1,2:N) = L(1:N-1,2:N) +...
 diag(ones(N-1,1));
\end{lstlisting}
\end{small}

\item GSVD.
\begin{lstlisting}[style = Matlab-editor]
[U,sm,X,V,W] = cgsvd(K,L).
\end{lstlisting}
\item Denoise TL signal. The temperature and the corresponding TL intensity are stored in  the first and second column of a matrix \texttt{TL}. This data is subsequently denoised by wavelet methods.
\begin{lstlisting}[style = Matlab-editor]
TLt = TL(:,1)+273.15; %temperature 
TLint = TL(:,2); %TL intensity
TLint_den = wdenoise(TLint);%wavelet denoising.
\end{lstlisting}
The signal \texttt{TLint\char`_den} is then further denoised by the stationary wavelet transform (\textsc{swt}) as implemented in the Wavelet Analyzer \textsc{app} of \textsc{matlab}. In this process, the signal is first extended to the required length and then denoised by using the \texttt{haar} wavelet to 3 levels of denoising. The cleaned signal is saved as \texttt{TLintp}. After denoising, it is interpolated to the temperature vector by the spline method.
\begin{lstlisting}[style = Matlab-editor]
b = interp1(TLt,TLintp,t,'spline'); 
\end{lstlisting}
\item Choose $\lambda_{opt}$. The optimized regularization parameter $\lambda_{opt}$ can be found from the corner of the L-curve \cite{L_curveHasen}, which is obtained from the GSVD of the problem. This involves solving Eq. \ref{eqn_KLni} by a series of sampled parameters via the \texttt{l\char`_curve} function,
\begin{lstlisting}[style = Matlab-editor]
[lambda_opt,~,~,~] = l_curve(U,...
sm,b,'Tikh',L,V);
\end{lstlisting}
where the desired parameter $\lambda_{opt}$ is \texttt{lambda\char`_opt}.
\item Solve the non-negative solution $\hat{\mathbf{x}}$. Firstly, the constraint-free solution $\mathbf{x}_{\lambda}$ (\texttt{x\char`_lambda}) will be calculated by calling the \texttt{tikhonov} function. The solution \texttt{x\char`_lambda} will be then bound at zero. A non-negative constraint will be imposed by implementing a non-negative least squares problem (Eq. \ref{eqn_KLni}) with the bound \texttt{x\char`_lambda} being a starting solution. The non-negative solution is denoted as \texttt{xhat}.
\begin{lstlisting}[style = Matlab-editor]
[x_lambda,~,~] =  tikhonov(U,sm,...
X,b,lambda_opt);
bhat = [b;zeros(size(L(:,1)))]; 
xhat0 = x_lambda; xhat0(xhat0<0)=0.0; 
lowerbound = zeros(size(xhat0)); 
upperbound = ones(size(xhat0)).*...
max(xhat0)*1.2; 
opts.Algorithm = 'trust-region-reflective'; 
opts.SubproblemAlgorithm = 'factorization'; % 
Ahat = [K;lambda_opt*L];
xhat = lsqlin(Ahat,bhat,[],[],[],...
[],lowerbound,upperbound,xhat0,opts);
%constrained linear least square
\end{lstlisting}
\end{itemize}
 
\section{Extracting the envelope}
\begin{figure}
\begin{center}
\includegraphics[width = 0.98\linewidth]{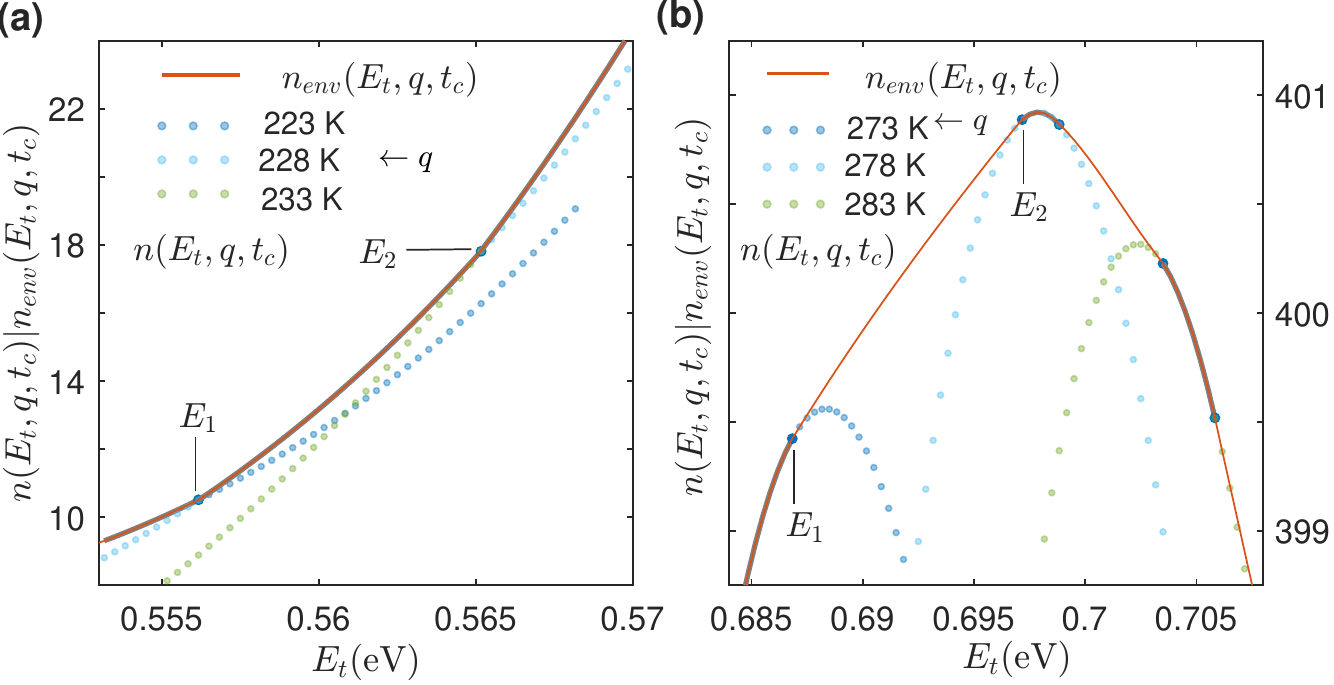} 
\caption{\textbf{Constructing the envelope $n_{env}(E_t,q,t_c)$}. \textbf{(a)} In certain $E_t$ ranges, the envelope $n_{env}(E_t,q,t_c)$ can be fully constructed by taking parts of data from $n(E_t,q,t_c)$ with consecutive $T_{ch}$. \textbf{(b)} In some $E_t$ ranges, interpolation is applied to calculate $n_{env}(E_t,q,t_c)$ when data points from $n(E_t,q,t_c)$ do not yield satisfactory results.   The gray thick lines outline the intersection of $n_{env}(E_t,q,t_c)$ and $n(E_t,q,t_c)$ with the endpoint shown in deep blue dots.}
\label{fig_extracEnvelope} 
\end{center}
\end{figure}
The envelope $n_{env}(E_t,q,t_c)$ is obtained from the entire set of electron population functions $n(E_t,q,t_c)$, corresponding to the different charging temperatures $T_{ch}$. In some $E_t$ range, the envelope $n_{env}(E_t,q,t_c)$ can be directly taken from the electron population function $n(E_t,q,t_c)$, as illustrated in Fig. \ref{fig_extracEnvelope}a. The electron population function $n(E_t,q,t_c)$ intersects with $n(E_t,q_1,t_c)$ and  $n(E_t,q_2,t_c)$ at $E_1$ and $E_2$, respectively, where
\begin{equation*}
q\rightarrow \text{228 K},\ q_1\rightarrow \text{223 K},\ q_2\rightarrow \text{233 K}.
\end{equation*}
Thus, $n_{env}(E_t,q,t_c)\approx n(E_t,q,t_c)$ in the range $(E_1,E_2)$ and this approximation is highlighted by a gray line. However, this method  may lead to unwanted artifacts. For example in Fig. \ref{fig_extracEnvelope}b, the electron population function $n(E_t,q,t_c)$ is expected to have its maximum between 339.5 and 401, which is higher than the value extracted by the aforementioned method. Hence,  no data points were taken from electron population functions in the indicated $(E_1,E_2)$ range.
All extracted data points are subsequently interpolated in the full range $E_t\in(E_a,E_b)$ by the method proposed by H. Akima \cite{interp1Akima}. It is clear that the interpolation also provides acceptable results in the range $(E_1,E_2)$ where no data points were taken from $n(E_t,q,t_c)$. 


\section{Trapping and recombination coefficients}
\begin{figure}
\centering
\includegraphics[width = 0.86\linewidth]{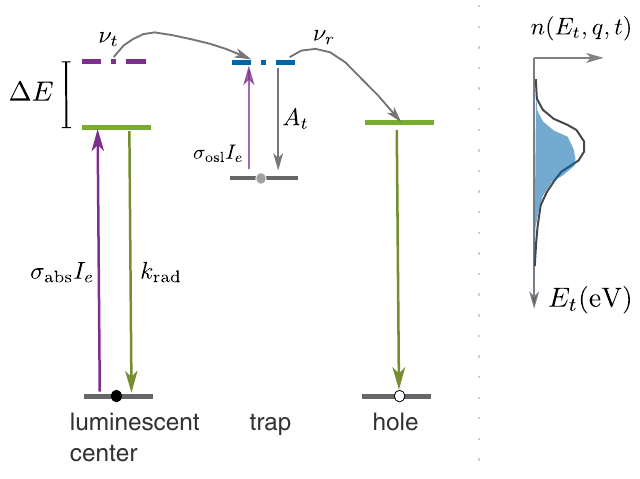} 
\caption{\textbf{Trapping and recombination coefficients}. The elementary "chemical reaction" of trapping and recombination can be described by Eq. \ref{eqn_chemReaction}, and the trapping and recombination coefficients can be approximated by analyzing the relative value of the coefficients depicted in the figure. For the luminescent activator, the non-radiative decay coefficient and the stimulated emission cross-section of the excited state have been neglected.}
\label{fig_activatortrasphole} 
\end{figure}

A luminescent activator (Fig. \ref{fig_activatortrasphole}) can be represented by its ground state and the excited state. This is also true for a trap. Within the framework of isolated-pair approximation, trapping and recombination only takes place within independent pairs and retrapping has been completely ignored. Here, we consider an elementary event of trapping and recombination. The kinetics of macroscopic densities of the pairs can be described by differential equations by applying the mean-field mass-action law \cite{DiffContReactbook}. Hence, the kinetics of elementary events  of trapping and recombination can be represented by a chain of mono-molecular "chemical reaction",
\begin{equation}
\ce{G <=>[$k_1$][$k_\text{-1}$]E ->[$k_2$]\phi}  \label{eqn_chemReaction}
\end{equation}
in which G and E represent the ground and excited state, respectively. Herein, $\phi$ represents the final 'product', which is an electron trapped in traps for trapping and a recombination of electron with hole for recombination, respectively. The density of G and E are often denoted as $[G]$ and $[E]$, respectively. The elementary trapping rate or recombination rate can be written as, 
\begin{equation}
R([E]) = k_2[E],
\end{equation}
in which $[E]$ represents the density of electrons at the excited states of luminescent activator for trapping or that of traps for recombination. 

According to Eq. \ref{eqn_chemReaction}, the following differential equations can apply \cite{QSA_QEA},
\begin{equation}
\frac{d}{dt}
\begin{bmatrix}
[G]\\
[E]
\end{bmatrix} = 
\begin{bmatrix}
 -k_1 & k_\text{-1}\\
  k_1 & -(k_\text{-1}+k_2)
\end{bmatrix} 
\begin{bmatrix}
[G]\\
[E]
\end{bmatrix}. \label{eqn_ODEeqn}
\end{equation}
This equation can be solved by matrix methods with initial condition $[G](t=0)=[G]_0,[E](t=0)=0$ \cite{QSA_QEA},
\begin{equation}
\begin{bmatrix}
[G]\\
[E]
\end{bmatrix} = \frac{[G]_0}{\lambda_1-\lambda_2}
\begin{bmatrix}
 k_1-\lambda_2 & -(k_1-\lambda_1)\\
 -k_1 & k_1
\end{bmatrix}
\begin{bmatrix}
\text{exp}(-\lambda_1 t)\\
\text{exp}(-\lambda_2 t)
\end{bmatrix} \label{eqn_ODEsol}
\end{equation}
in which $\lambda_1$ and $\lambda_2$ are the eigenvalues of the coefficients matrix in Eq. \ref{eqn_ODEeqn}, 
\begin{subequations}
\begin{align}
\lambda_1 = & \frac{1}{2}\left[k_1+k_\text{-1}+k_2+\sqrt{(k_1+k_\text{-1}+k_2)^2-4k_1k_2}\right]\\
\lambda_2 = & \frac{1}{2}\left[k_1+k_\text{-1}+k_2-\sqrt{(k_1+k_\text{-1}+k_2)^2-4k_1k_2}\right]
\end{align}
\end{subequations}
The time for $[E]$ to reach its maximal value is thus \cite{QSA_QEA},
\begin{equation}
\tau_m = \frac{1}{\lambda_1-\lambda_2}\text{ln}\left(\frac{\lambda_1}{\lambda_2}\right)
\end{equation}
and when $t>\tau_m$ it is safe to set $\text{exp}(-\lambda_1t)$ to zero in Eq. \ref{eqn_ODEsol}, leading to approximate solutions
\begin{subequations}
\begin{align}
[G] \approx & \frac{(\lambda_1-k_1)[G]_0}{\lambda_1-\lambda_2}\text{exp}(-\lambda_2t), \\
[E] \approx & \frac{k_1[G]_0}{\lambda_1-\lambda_2}\text{exp}(-\lambda_2t).
\end{align}
\end{subequations}

In the context of TL or PersL, the density $[G]+[E]$ is more convenient i.e., before it recombines with a hole, an electron is reckoned as being trapped at either the ground or excited state of the trap, whose density can be described by the electron population function $n(E_t,q,t_c)$. Therefore, the following ratio is calculated:
\begin{equation}
r([E]) = \frac{[E]}{[G]+[E]}\approx \frac{k_1}{\lambda_1}. \label{eqn_ratiorE}
\end{equation}

\subsection{Trapping coefficient}
We first consider the case of trapping. The presence of a thermal barrier makes the trapping coefficient to follow the Arrhenius relation with activation energy $\Delta E$. The coefficients for Eq. \ref{eqn_chemReaction} are
\begin{subequations} 
\begin{align}
k_1 &=  \sigma_\text{abs}I_e(\lambda),\\
k_\text{-1} &=  k_\text{rad}, \\
k_2 & =  \nu_t\text{exp}\left(-\frac{\Delta E}{k_\textsc{b}T}\right),
\end{align}
\end{subequations}
in which stimulated emission of the luminescent activator has been neglected because optical excitation in the experiments is weak. It can be seen that,
\begin{equation}
k_\text{-1} \gg k_1 \quad \mathrm{and} \quad k_\text{-1} \gg k_2,
\end{equation}
which leads to an approximation for $r([E])$,
\begin{equation}
r([E])\approx \frac{k_1}{k_\text{-1}}
\end{equation}
and thus the trapping rate can be calculated by 
\begin{eqnarray}
R([E]) & =& k_2\times\left([G]+[E]\right)\times r([E]) \nonumber\\
&=& \frac{k_1k_2}{k_\text{-1}}\times\left([G]+[E]\right).
\end{eqnarray}
It is quite clear now that the trapping coefficient turns out to be,
\begin{equation}
\boxed{
k_\text{trap}(\Delta E,q) = \frac{\nu_t\sigma_\text{abs}I_e(\lambda)}{k_\text{rad}}\textrm{exp}\left(-\frac{\Delta E}{k_\textsc{b}T_{ch}}\right)
},
\end{equation}
which is Eq. 11 in the paper. \\
 
\subsection{Recombination coefficient}
The non-radiative decay rate coefficient for an excited trap is given by $A_t$. The thermal detrapping rate coefficient is $A_t\text{exp}\left(-\frac{E_t}{k_\textsc{b}T}\right)$ with trap depth $E_t$. Therefore, the coefficients for the "chemical reaction", Eq. \ref{eqn_chemReaction} now becomes:
\begin{subequations} 
\begin{align}
k_1 &=  \sigma_\text{osl}I_e(\lambda)+A_t\text{exp}\left(-\frac{E_t}{k_\textsc{b}T}\right),\\
k_\text{-1} &=  A_t, \\
k_2 & =  \nu_r.
\end{align}
\end{subequations}
In this case, the relation holds,
\begin{equation}
k_1\ll k_\text{-1} \quad \mathrm{and} \quad k_1\ll k_2.
\end{equation}
which leads to an approximation,
\begin{equation}
r\left([E]\right)\approx \frac{k_1}{k_\text{-1}+k_2}.
\end{equation}
The rate of recombination becomes
\begin{align}
R([E]) & = k_2\times\left([G]+[E]\right)\times r([E])\\
& = \frac{k_1k_2}{k_\text{-1}+k_2}\times\left([G]+[E]\right),
\end{align}
which yields the coefficient of recombination,
\begin{equation}
\boxed{
k_\text{rcb}(E_t,q) = \frac{A_t}{A_t+\nu_r}\left[\nu_r \textrm{exp}\left(-\frac{E_t}{k_\textsc{b}T}\right) +\frac{\nu_r}{A_t}\sigma_\text{osl}I_e(\lambda)\right]
},
\end{equation}
which is Eq. 12 in the paper.

\section{Dose dependency}
For first-order kinetics, current belief in TL community is that the shape of the electron population function (or TL glow curve) is independent of charging duration with fixed charging irradiance $I_e(\lambda)$, i.e. the dose. On the contrary, a shift of the TL glow curve is expected with increasing dose for non-first-order kinetics \cite{e_trapmechanism_ProcPhysSoc}. However, these predictions are based on the assumption that only one discrete trap depth is present in the material under study. In the current paper, a distribution of trap depths has been assumed, and therefore the electron population function will be investigated as a function of charging duration $t_{ch}$ at fixed charging irradiance $I_e(\lambda)$.  
\begin{figure}
\centering
\includegraphics[width = 0.98\linewidth]{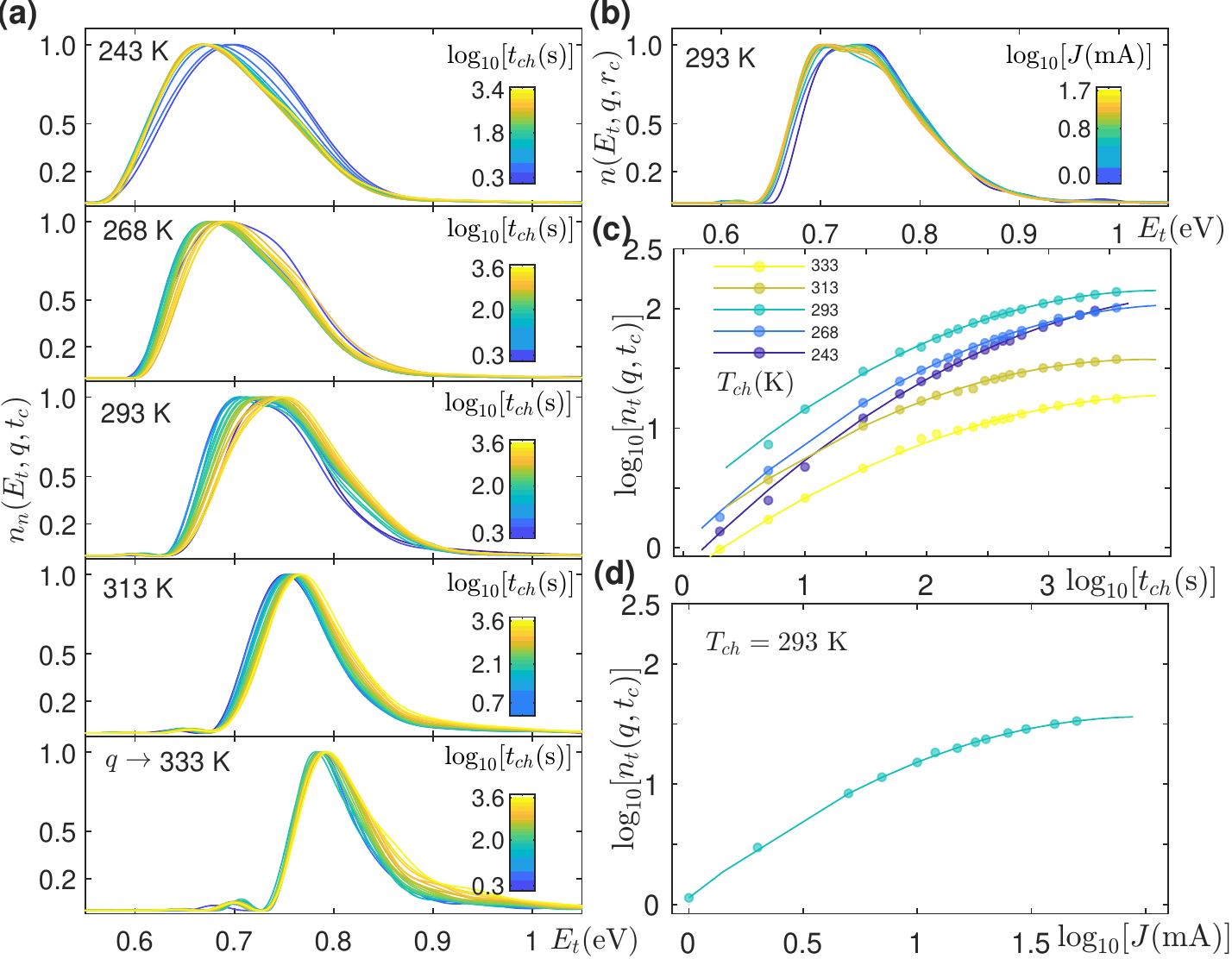} 
\caption{\textbf{Dose dependency}. \textbf{(a)} The trap depth at the maximum of the electron population function, $E_m$, increases with increasing charging duration, $t_{ch}$, illustrated for various charging temperatures $T_{ch}$. \textbf{(b)} The shape of the electron population function changes only marginally with increasing LED driving current ($J=$ 1~mA - 50~mA). \textbf{(c)} $\text{log}_{10}\left[n_t(E_t,q,t_c)\right]$ is a quadratic function of $\text{log}_{10}[t_{ch}(\text{s})]$, illustrated for various $T_{ch}$. \textbf{(d)} $\text{log}_{10}\left[n_t(E_t,q,t_c)\right]$ is a quadratic function of $\text{log}_{10}[J(\text{mA})]$, illustrated for $T_{ch}=$ 293~K and $t_{ch}=$ 30~s.} 
\label{fig_epopShapetch} 
\end{figure} 

The normalized electron population function,
\begin{equation*}
n_n(E_t,q,t_c)  = \frac{n(E_t,q,t_c)}{\underset{E_t }{\mathrm{max}\;} n(E_t,q,t_c)}
\end{equation*}
reveals the shapes of $n(E_t,q,t_c)$ at different charging conditions $q$. For every charging temperature ($T_{ch}=$ 243, 268, 293, 313, and 333 K), the trap depth corresponding to the maximum of $n_n(E_t,q,t_c)$, i.e. $E_m$, shifts to higher values with increasing charging duration $t_{ch}$ (Fig. \ref{fig_epopShapetch}a) before stabilization. However, the shape of $n_n(E_t,q,t_c)$ remains more or less unchanged with increasing LED current (1 mA - 50 mA) for a fixed charging duration ($t_{ch}=30~\mathrm{s}$) (Fig. \ref{fig_epopShapetch}b). The total number of trapped electrons per volume, 
\begin{equation}
	n_t(q,t_c) = \int_0^\infty n(E_t,q,t_c)~\mathrm{d}E_t,
\end{equation}	
is related to the charging duration $t_{ch}$ or driving current of the LED $J$ by a quadratic function in the log-log scale,
\begin{equation}
y=ax^2+bx+c \label{eqn_parabola}
\end{equation}
in which $y = \text{log}_{10}\left[n_t(q,t_c)\right]$,  $x=\text{log}_{10}(t_{ch})$ or $x=\text{log}_{10}(J)$ (Fig. \ref{fig_epopShapetch}c-d).

\begin{figure}
\centering
\includegraphics[width = 0.98\linewidth]{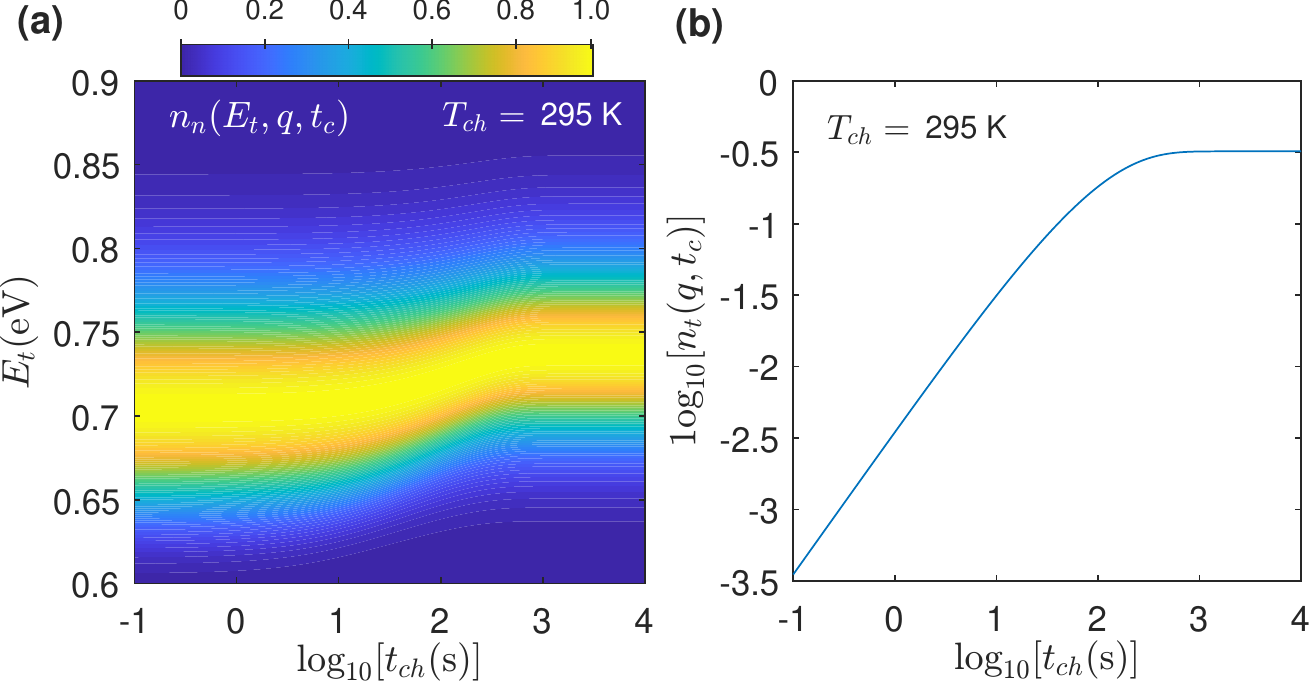} 
\caption{\textbf{Simulation of charging kinetics}. \textbf{(a)} The color plot of $n_n(E_t,q,t_c)$ for various $t_{ch}$ shows that the trap depth at the $n_n(E_t,q,t_c)$ maximum, i.e. $E_m$, starts shifting to higher values (at  $\sim 10$ s) and stabilizes for high $t_{ch}$ (at $\sim 10^3$ s ). \textbf{(b)} The total number of trapped electrons per volume $n_t(q,t_c)$ is a function of charging duration $t_{ch}$, following Eq. \ref{eqn_parabola} when $t_{ch}<10^3\text{ s}$.}
\label{fig_tchsim295K} 
\end{figure}

To understand these observations, the electron population is simulated for a Gaussian distribution of trap depth,
\begin{equation*}
N(E_t)=\frac{1}{\sqrt{2\pi \sigma^2}}\text{exp}\left[-\frac{(E_t-E_u)^2}{2\sigma^2}\right],
\end{equation*}
with $E_u=$ 0.70~eV and $\sigma=$ 0.05~eV. These parameters were arbitrarily chosen to resemble the derived trap depth distribution for BaSi$_2$O$_2$N$_2$:Eu$^{2+}$. The simulation of $n_n(E_t,q,t_c)$ reveals that $E_m$ increases almost linearly with increasing $\text{log}_{10}(t_{ch})$ (Fig. \ref{fig_tchsim295K}a). In addition, the simulated $\text{log}_{10}\left[n_t(q,t_c)\right]$ can be related to $\text{log}_{10}(t_{ch})$ by a quadratic function when $t_{ch}<\sim 10^3 \text{ s}$ (Fig. \ref{fig_tchsim295K}b). It is interesting to note that the shift of the electron population maximum with increasing charging duration can be explained by first-order kinetics with trap depth distribution rather than non-first order kinetics. \par

Herein, we provide the \textsc{matlab} code for the simulation.
\begin{itemize}
\item
\begin{lstlisting}[style = Matlab-editor]
% the initialization
kB = 0.08617; % meV/K
nu_r = 1e10; nu_t = nu_r; A_t = 1e12;
sigma_osl = 1e-17; sigma_abs = 3e-18; % cross-sections
krad = 1.54e6; % radiative rate of Eu in BaSiON, Hz
dE = 255; Ie_exp = 5e15; beta = 0.5; 

syms deltaE Ie t T Et Ti
ktrap = @(deltaE,T,Ie) 1*nu_t*exp(-deltaE./(kB*T))*sigma_abs.*Ie/krad;
krcb = @(Et,T,Ie) (nu_r*exp(-Et./(kB*T))+sigma_osl.*Ie*nu_r/A_t)/(1+nu_r/A_t);
filling = @(Et,deltaE,T,Ie,t) ktrap(deltaE,T,Ie)./(ktrap(deltaE,T,Ie)+krcb(Et,T,Ie)).*(1-exp(-(ktrap(deltaE,T,Ie)+krcb(Et,T,Ie)).*t));
% the above is the filling function
tInt = @(Et,T) 0.25*kB*nu_r/beta*(T.^2./Et).*exp(-Et./T/kB)./sqrt(1+4*kB*T./Et); % effective cooling/heating rate is 4beta
TempInt = @(Et,T) tInt(Et,T)-tInt(Et,T-30); % cooling down to T0 = T-30 K
cool = @(Et,T) exp(-TempInt(Et,T));

tch = logspace(-3.0,4,200)'; % tch in log10 space
dE = 1; Ea = 50; Eb = 2e3; E = (Ea:dE:Eb)'; % meV
[tgrd,Egrd] = meshgrid(tch,E);
Tch = 295;
Fill = filling(Egrd,dE,Tch,Ie_exp,tgrd);
Fillcool = Fill.*cool(Egrd,Tch);
% calculate gaussian distribution
Eav = 700; Estd = 50; % 
NE = 1/sqrt(2*pi*Estd^2)*exp(-0.5*(E-Eav).^2./Estd^2); % gaussian distribution;
Nn = (NE.*Fillcool)./max(NE.*Fillcool,[],1);% normalized electron population function
lgnt = log(sum(NE.*Fillcool)*dE)/log(10); % n_t in log10 scale
\end{lstlisting}
\end{itemize}

\section{Simplification of the kernel}
The temperature integral $F(E_t,T_0)$ is significantly smaller than $F(E_t,T)$ when $T_0$ is about $\sim$ 30 K below $T$, enabling a simplification of the kernel, given by Eq. 24 in the paper. Furthermore, the PersL decay profile can be analyzed by the Fredholm integral using kernel Eq. 29 in the paper. The details of the derivation will be given in the following. \par

At a given temperature $T>T_0+20\sim 30$ K, the kernel $K(E_t,T)$  can be approximated by neglecting $F(E_t,T_0)$:
\begin{linenomath*}
\begin{equation}
K(E_t,T)=\text{exp}\left[-\frac{E_t}{k_\textsc{b}T}-F(E_t,T)\right]. \label{eqn_KernelnoFET0}
\end{equation}
\end{linenomath*}
The derivative of $\text{ln}\left[K(E_t,T)\right]$ with respect to $E_t$ is,
\begin{linenomath*}
\begin{equation}
\begin{split}
\frac{\partial}{\partial E_t}\textrm{ln}\left[K(E_t,T)\right]&=-\frac{1}{k_{\textsc{b}}T}+\\
&\frac{\nu_r}{\beta}\int_0^T\frac{1}{k_{\textsc{b}}T'}\textrm{exp}\left(-\frac{E_t}{k_{\textsc{b}}T'}\right)dT'. \
\end{split}
\label{eqnAPP_derlnKernelEtT}
\end{equation}
\end{linenomath*}
After following substitution,
\begin{linenomath*}
\begin{equation*}
t= \frac{E_t}{k_{\textsc{b}}T'}, \qquad {x= \frac{E_t}{k_{\textsc{b}}T}},
\end{equation*}
\end{linenomath*}
the second term in Eq. \ref{eqnAPP_derlnKernelEtT} can be simplified as,
\begin{linenomath*}
\begin{align*}
\frac{\nu_r}{\beta}\int_0^{T}\frac{1}{k_{\textsc{b}}T'}\text{exp}\left(-\frac{E_t}{k_{\textsc{b}}T'}\right)dT'&=\frac{\nu_r}{k_{\textsc{b}}\beta}\int_{x}^{\infty}\frac{\text{exp}\left(-t\right)}{t}dt \nonumber \\ 
&=\frac{\nu_r}{k_{\textsc{b}}\beta}E_1\left(x\right),
\end{align*}
\end{linenomath*}
where the exponential integral $E_1(x)$ can be approximated by \cite{expintRef},
\begin{linenomath*}
\begin{equation}
E_1\left(x\right)=\frac{\text{exp}(-x)}{x+1}. \label{eqnApp_E1Approx}
\end{equation}
\end{linenomath*}
Setting Eq. \ref{eqnAPP_derlnKernelEtT} to zero means,
\begin{linenomath*}
\begin{equation}
-\frac{1}{k_{\textsc{b}}T}+\frac{\nu_r}{k_{\textsc{b}}\beta}\frac{\text{exp}\left(-x\right)}{x+1} = 0, \label{eqnApp_deriveq0}
\end{equation}
\end{linenomath*}
which leads to the root,
\begin{linenomath*}
\begin{equation}
x_s = W(\text{e}\nu_r T/\beta)-1 \label{eqnApp_xroot}
\end{equation}
\end{linenomath*}
where $\text{e}=\text{exp}(1)$ and $W(x)$ is the Lambert $W$ function with branch $n=0$. This results in Eq. 25b in Sec. IVA of the paper,
\begin{equation}
\boxed{
E_s(T) = k_\textsc{b}T\left[W(\text{e}\nu_r T/\beta)-1\right]}.
\end{equation}
According to Eq. \ref{eqnApp_deriveq0}, the following holds,
\begin{linenomath*}
\begin{equation*}
\text{exp}(-x_s)= \frac{\beta(x_s+1)}{\nu_rT}, \label{eqnApp_expxs}
\end{equation*}
\end{linenomath*}
which further leads to,
\begin{linenomath*}
\begin{equation}
\text{exp}\left(-\frac{E_s}{k_{\textsc{b}}T}\right)\frac{\nu_r}{\beta}\frac{k_{\textsc{b}}T^2}{E_t}=\frac{E_s+k_{\textsc{b}}T}{E_t} \label{eqnApp_expxs_xsRelation}
\end{equation}
\end{linenomath*}
The kernel Eq. \ref{eqn_KernelnoFET0} was initially expressed as
\begin{linenomath*}
\begin{equation}
\begin{split}
K\left(E_t,E_s\right) =& \frac{\nu_r}{\beta}\text{exp}\Bigl(-\frac{E_s}{k_{\textsc{b}}T}\Bigr)\text{exp}\Bigl[-\frac{E_t-E_s}{k_{\textsc{b}}T}\\
&-\text{exp}\Bigl(-\frac{E_t-E_s}{k_{\textsc{b}}T}\Bigr)\text{exp}\Bigl(-\frac{E_s}{k_{\textsc{b}}T}\Bigr)\\
&\times\frac{\nu_r}{\beta}\frac{k_{\textsc{b}}T^2}{E_t}\frac{1}{\sqrt{1+4k_{\textsc{b}}T/E_t}}\Bigr]
\end{split}
\label{eqn_KernelKEtEs}
\end{equation}
\end{linenomath*}
which results in Eq. 25a immediately by utilizing Eq. \ref{eqnApp_expxs_xsRelation}.  The magnitude of the kernel Eq.~\ref{eqn_KernelKEtEs} is,
\begin{linenomath*}
\begin{equation}
\boxed{
\frac{\nu_r}{\beta}\text{exp}\left(-\frac{E_s}{k_{\textsc{b}}T}\right)=W(\text{e}\nu_rT/\beta)/T}\label{eqnApp_magnitude}
\end{equation}
\end{linenomath*}
which changes with $T$ at the rate of 
\begin{linenomath*}
\begin{equation}
\frac{\partial}{\partial T}\frac{\nu_r}{\beta}\text{exp}\left(-\frac{E_s}{k_{\textsc{b}}T}\right)=-\frac{W(\text{e}\nu_rT/\beta)}{T^2}\frac{W(\text{e}\nu_rT/\beta)}{W(\text{e}\nu_rT/\beta)+1}.
\end{equation}
\end{linenomath*}
The derivative of $E_s$ with respect to $T$ is
\begin{linenomath*}
\begin{equation}
\frac{\partial E_s}{\partial T} = k_{\textsc{b}}W(\text{e}\nu_rT/\beta)\frac{W(\text{e}\nu_rT/\beta)+2}{W(\text{e}\nu_rT/\beta)+1}-k_{\textsc{b}} \label{eqnApp_Esdriv2T}
\end{equation} 
\end{linenomath*}

Similarly, the decay profile of persistent luminescence can be given in the following integral equation,
\begin{equation}
I(t_0) = \int_0^{\infty}K(E_t,t_0)n(E_t,q,t=0)dE_t
\end{equation} 
in which the kernel reads,
\begin{equation}
K(E_t,t_0)=\nu_r\text{exp}\Bigl(-\frac{E_t}{k_\textsc{b}T}\Bigr)\text{exp}\Bigl[-\int_0^{t_0}\nu_r\text{exp}\Bigl(-\frac{E_t}{k_\textsc{b}T}\Bigr)dt\Bigr].
\label{eqn_KEt0PersLdef}
\end{equation}
It can be rewritten in the form,
\begin{equation*}
\begin{split}
K(E_t,t_0)=&\nu_r\text{exp}\Bigl(-\frac{E_s(t_0)}{k_\textsc{b}T}\Bigr)\times\\
&\text{exp}\Bigl[-\frac{E_t-E_s(t_0)}{k_\textsc{b}T}\\
&-\text{exp}\Bigl(-\frac{E_t-E_s(t_0)}{k_\textsc{b}T}\Bigr)\times \nu_r t_0\text{exp}\Bigl(-\frac{E_s(t_0)}{k_\textsc{b}T}\Bigr)\Bigr],
\end{split}
\end{equation*}
from which the characteristic trap depth $E_s(t_0)$ can be found by seeking the root of 
\begin{equation*}
\nu_rt_0\text{exp}\left(-\frac{E_s(t_0)}{k_\textsc{b}T}\right) - 1 = 0
\end{equation*}
leading to 
\begin{equation}
\boxed{
E_s(t_0) = k_\textsc{b}T\text{ln}(\nu_rt_0)},
\end{equation}
which is Eq. 28b in the paper.
The magnitude of the kernel $K(E_t,t_0)$ is,
\begin{equation*}
\nu_r\text{exp}\left(-\frac{E_s(t_0)}{k_\textsc{b}T}\right)=\frac{1}{t_0}
\end{equation*}
and thus the kernel can be written in the compact form,
\begin{equation}
\boxed{
K(E_t,t_0)=\frac{1}{t_0}\text{exp}\left[-\frac{E_t-E_s(t_0)}{k_\textsc{b}T}-\text{exp}\Bigl(-\frac{E_t-E_s(t_0)}{k_\textsc{b}T}\Bigr)\right]},
\label{eqn_KEt0PersLcompact}
\end{equation}
which is Eq. 28a in the paper.

%
%


\end{document}